\documentclass[aps,prl,10pt,twocolumn,showpacs,superscriptaddress,amsmath,amssymb]{revtex4-1}

\usepackage{graphicx}
\usepackage[normalem]{ulem}
\usepackage[usenames,dvipsnames]{color}
\usepackage[nointegrals]{wasysym}
\usepackage{siunitx}
\usepackage{hyperref}
\usepackage{bbm}
\usepackage{commands} 
\usepackage{cleveref}

\pacs{
73.23.Hk, 
73.63.-b, 
73.63.Kv, 
}

\begin{document}

\title{Lissajous rocking ratchet}

\author{Sergey Platonov} \affiliationMunich
\author{Bernd K\"astner} \affiliationPTB
\author{Hans W. Schumacher} \affiliationPTB
\author{Sigmund Kohler} \affiliationMadrid
\author{Stefan Ludwig}
\thanks{Present address: Paul-Drude-Insti\-tut f\"ur Fest\-k\"or\-per\-elek\-tro\-nik, Haus\-vog\-tei\-platz 5--7, 10117 Berlin, Germany}
\affiliationMunich

\date{\today}

\pacs{%
73.63.Kv, 
05.60.-k, 
85.35.Gv 
}

\begin{abstract}
Breaking time-reversal symmetry (TRS) in the absence of a net bias can give rise to directed steady-state non-equilibrium transport phenomena such as ratchet effects. Here we present, theoretically and experimentally, the concept of a Lissajous rocking ratchet as an instrument based on breaking TRS. Our system is a semiconductor quantum dot (QD) with periodically modulated dot-lead tunnel barriers. Broken TRS gives rise to single electron tunneling current. Its direction is fully controlled by exploring frequency and phase relations between the two barrier modulations. The concept of Lissajous ratchets can be realized in a large variety of different systems, including nano-electrical, nano-electromechanical or superconducting circuits. It promises applications based on a detailed on-chip comparison of radio-frequency signals.
\end{abstract}

\maketitle

Ratchets cause directed particle motion due to a combination of broken symmetry and non-equilibrium forces, where the latter may be deterministic or fluctuating. The most famous example is Feynman's \textit{flashing ratchet} which uses a pulsating spatially asymmetric potential to actively turn fluctuations into work \cite{Feynman1963a,Reimann2002a,Khrapai2006a,Hanggi2009a}. Another species is the \textit{rocking ratchet} driven by forces periodic in time but with broken spatio-temporal symmetry.  A simple example of a rocking ratchet is a pump which transports electrons one-by-one through a QD driven by two external periodic forces with a relative phase breaking the symmetry \cite{Ono2003,Jehl2013}. This is in contrast to the somewhat simpler turnstile where the spatial symmetry is broken by a finite dc voltage \cite{Kouwenhoven1991a}. In the non-adiabatic limit electron pumps are investigated for their suitability as current standard \cite{Blumenthal2007a,Kaestner2008b}.

\begin{figure*}[ht]
\includegraphics[width=\textwidth]{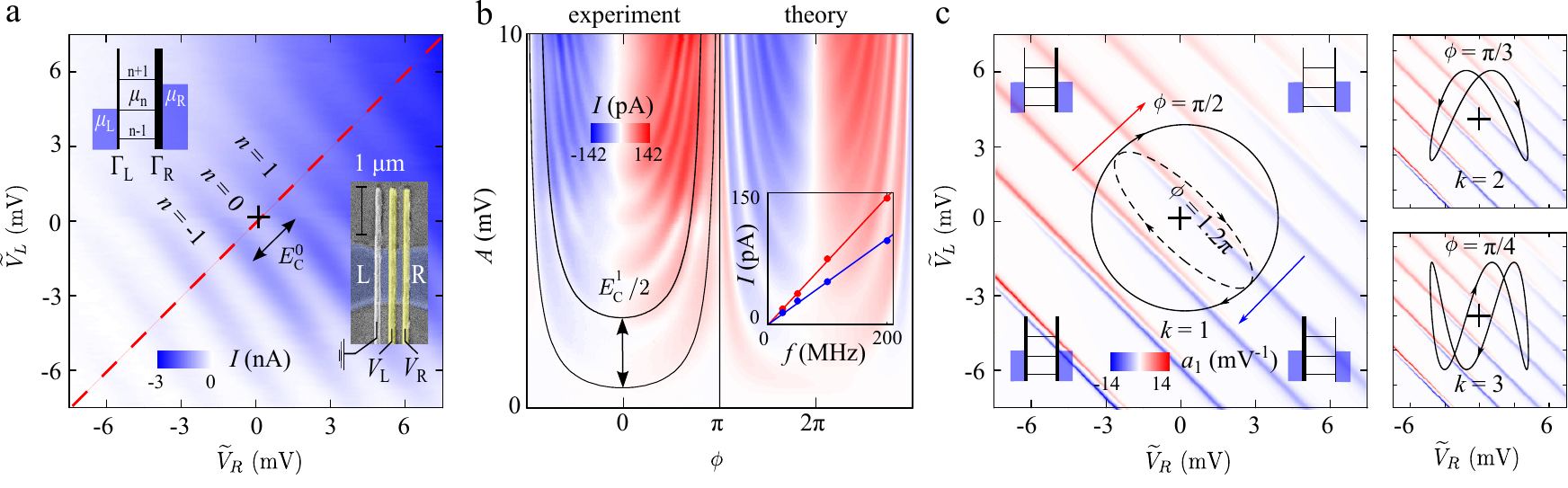}
\vspace{-5mm}
\caption{\label{fig1}
(a) Current $I$ through the QD as function of gate voltages $\tvlr$ in response to a dc voltage $V=(\mur-\mul)/e=100\,\mu$V. A double arrow indicates the charging energy $\ec^0$ at the working point (black cross). Lower inset: scanning electron microscope image of the sample. The wafer is etched at the gray area, only the blue channel contains 2DES. Two Ti/Au gates (yellow) are used to define a QD, a third gate (gray) is electrically grounded. Upper inset: QD sketch; vertical lines are electrostatic barriers with tunnel couplings $\glr$ controlled by gate voltages. Horizontal lines are chemical potentials $\mun$ of the QD, blue areas indicate occupied states in the degenerate leads at chemical potentials $\mulr$. 
(b) $\tvlr$ are modulated according to \eq{lissajous}, both with frequency $f=200\,$MHz ($k=1$). Measured (left) versus calculated (right, $V=1\,\mu$V) current $I$ versus modulation amplitude and phase. Inset: measured $I(f)$ at $A=9$\,mV for the two phases $\phi\simeq0.3\pi$ (blue), $\phi\simeq0.6\pi$ (red).
(c) Vector field $\vec a$ (explained in text and \suppl) versus gate voltages and typical Lissajous trajectories for $k=1,2,3$ according to \eq{lissajous}. Black crosses mark the working point, QD sketches indicate relative strengths of tunnel couplings to the leads.
}
\end{figure*}

In this article we restrict ourselves to the adiabatic regime and study a
generic implementation of a rocking ratchet by applying two time-periodic
forces, phase-locked at various commensurate frequencies. In our implementation we measure the dc current $I$ through a QD embedded in the two-dimensional electron system (2DES) 90\,nm beneath the surface of an etched  1\,$\mu$m wide channel of GaAs/AlGaAs heterostructure. The 2DES is cooled to $\sim$100\,mK where its carrier density is $n_e\simeq2.83\times10^{15}$\,m$^{-2}$ and its mobility is $\mu_e\simeq320\,\mathrm{m}^2\,\mathrm{V}^{-1}\,\mathrm{s}^{-1}$. We control the QD by applying voltages to two metal gates [see lower inset in \fig{fig1}{a}].

For a first orientation we present in \fig{fig1}{a} a stability diagram of
our QD measured at finite dc voltage $V=(\mur-\mul)/e=100\,\mu$V applied
between its two leads (at chemical potentials $\mulr$). Plotted is the
current $I$ as a function of gate voltages $\tvl$ and $\tvr$ applied to the
left (L) versus right (R) gate [yellow in the rhs inset of \fig{fig1}{a}].
The axes $\tvlr$ are offset relative to the actually applied voltages,
such that $\tvl=\tvr=0$ at our working point, which is marked in
\fig{fig1}{a} by a black cross. We define the chemical potential $\mun$ of
the QD as the energy needed to add the next electron to it, where $n$ is an
index number and we choose $n=0$ for the dot level closest to the working
point.  For $\mur\gtrsim\mun\gtrsim\mul$ the Coulomb blockade is lifted and
a current $I$ flows along broadened lines oriented
perpendicular to the bisecting line (dashed line). With increasing $\tvlr$,
the barriers coupling the QD to both leads shrink causing $I$ to grow.  At
the bisecting line, the QD is symmetrically coupled to both leads but with
increasing distance, one barrier shrinks and the other one grows, causing
$I$ to decrease.

We parametrize the QD by its dot-lead tunnel rates $\glr$ and its charging
energies $\ec(n)=\mun-\mu_{n-1}$, where at our working point,
$\ec(0)\equiv\ec^0\simeq1.27\,$meV.  Exploring a limited range of the stability diagram the capacitive coupling between gates and
the QD can be modeled by the linear relation $\mun=\mun^0-\km(\tvl+\tvr)$
with $\km\simeq0.38e$. Here $e$ is the elementary charge and $\mun^0$ are
the chemical potentials for $\tvl=\tvr=0$. The tunnel couplings
$\gl(\tvl)$ and $\gr(\tvr)$ are controlled by the respective gate voltages
$\vlr$, where we neglect the very weak cross couplings $\gl(\tvr)$ and $\gr(\tvl)$.
The $\glr(\tvlr)$ are determined by fitting a theoretical prediction to the
current in \fig{fig1}{a}. Our calibrations are detailed
in the Supplemental Material, Ref.~\cite{supplement}. In the following we
define rocking ratchets by making use of the mutually coupled dependencies
of $\glr$ and $\mun$ on both gate voltages.

To break the TRS, we periodically modulate $\tvl(t)$ and $\tvr(t)$ such that they define trajectories along Lissajous figures centered at the working point:
\begin{equation}
\label{lissajous}
\vkphi
= \begin{pmatrix} \tvl(t)\\ \tvr(t) \end{pmatrix}
=A\begin{pmatrix} \cos(2\pi ft-\phi)\\ \cos(k2\pi ft) \end{pmatrix},
\end{equation}
where, for simplicity, we restrict ourselves to integer frequency ratios $k$. We prepare the two modulation amplitudes to be identical, $A$. The trajectories can cross several charging lines of the stability diagram.

Our first experiment resembles an adiabatic electron pump defined by \eq{lissajous} with $k=1$, \ie identical frequencies.  \Fig{fig1}{b} compares a typical measurement (lhs) with model predictions (rhs), both for $f=200\,$MHz and $\mul\simeq\mur$. The current, plotted as a function of $A$ and phase $\phi$, displays clear resonances following lines of constant $A\cos(\phi/2)$ (solid lines).  These resonances indicate discrete current contributions of the $n$-electron QD levels separated by $\km \Delta A=\ec(n)/2$, where the factor $2$ accounts for two gates being modulated.  During each pumping cycle approximately $\text{int}(2A\km/\ec^0)$ ground state levels fulfill $\mun(t)=\mulr$ twice.  For $f\le200\,$MHz we find $I\propto f$ at fixed $\phi$ and $A$ [inset in \fig{fig1}{b}] which confirms that we operate in the adiabatic regime, where the QD is always in its momentary ground state and the charge transferred across it per cycle is constant \cite{supplement}. Each QD level is unoccupied whenever $\mun(t)>\mulr$ and its $n$-electron ground state is occupied for $\mun(t)<\mulr$. Whenever $\mun(t)\simeq\mulr$ an electron tunnels into the QD if $\text d\mun/\text d t<0$ and out of the QD if $\text d\mun/\text d t>0$. With which lead the electron is thereby exchanged depends on $\phi$ and the ratio $\gl/\gr$, which is modulated in time, see~\eq{lissajous} and above.

This scenario resembles a rocking ratchet \cite{Hanggi2009a} where the current changes direction at $\phi=0,\pi$, independent of $k$. For $k=1$ it gives rise to the two-fold symmetry of a pump observed in \fig{fig1}{b}. The symmetry can be compromised for four reasons (independent of $k$): (i) dissipation by transitions within the QD's excitation spectrum. This, however, would go along with non-adiabaticity which we already excluded [inset in \fig{fig1}{b}]; (ii) a spatially asymmetric local disorder potential which influences the ratio $\gl/\gr$ as function of $n$, depending on the electronic probability distribution. This is ignored in our model but can explain deviations to the measured current, particularly near $\phi=0$ where the QD's chemical potentials $\mun(t)$ are modulated strongest and additional spatial disorder causes rectification \cite{0295-5075-44-3-341}; (iii) the $n$-dependence of the charging energy and the choice of the working point; (iv) a dc voltage between the leads ($V\ne0$). Points (iii) and (iv) are included in our model and discussed below.

A formal description of the general case for any integer $k$ including different frequencies modulating the two gates, is detailed in \suppl\ and can be summarized as follows: We start from an expression for the current through the time-periodic system obtained with Floquet transport theory \cite{Kohler2005a,Forster2014}.  Taylor expansion up to first order in $\Omega$ provides the adiabatic limit of the dc current: ~
\begin{equation}
\label{current}
I = \bar G V + Q^\mathrm{cycle} f + I^\text{rect},
\end{equation}
\begin{figure*}[t] 
\includegraphics[width=1\textwidth]{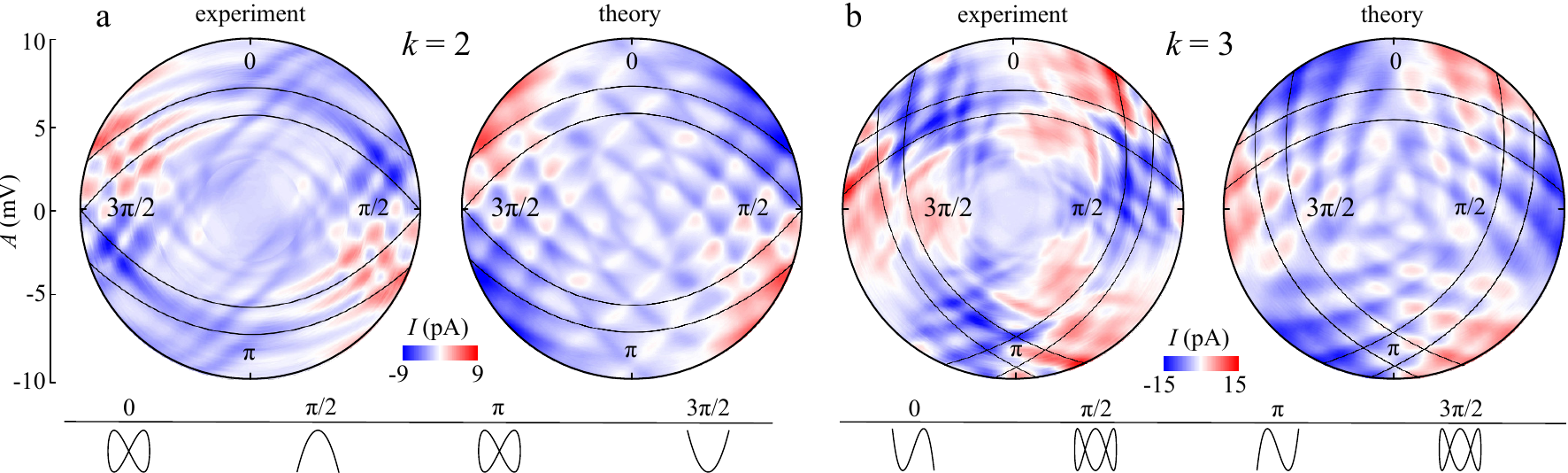}
\vspace{-5mm}
\caption{\label{fig2}
Measured (left) and computed (right) current as a function of modulation
amplitude $A$  and phase difference $\phi$ between $\vl(t)$  and $\vr(t)$:
(a) $\vl(t)$ at 50\,MHz and $\vr(t)$ at 100\,MHz ($k=2$), see
\eq{lissajous}, and (b) at 50\,MHz and 150\,MHz ($k=3$). Solid lines
indicate the resonances $\mun\simeq\mu_{L,R}$ for $n=3,4$.
Bottom: Lissajous figures for $k=2$ and $k=3$. 
}
\end{figure*}%
which can be traced back to the transport properties of the time-independent system for the parameters along the trajectory $\vkphi$, defined in \eq{lissajous}. The first term of \eq{current} contains the time-averaged conductance of the QD, $\bar G$. It can be understood as dc current which flows whenever any resonance fulfills the condition $\mul\lesssim\mun\lesssim\mur$ (or $\mur\lesssim\mun\lesssim\mul$). For the theory data in \fig{fig1}{b} we have used $V=1\,\mu$V which provides good agreement with the measured data. This tiny dc voltage is not purposely applied but caused by the current amplifier. 

The second term of \eq{current}  is the dc current caused by the pumped charge, $Q^\mathrm{cycle}$. This ratchet current can be expressed as a closed loop integral of a vector field $\vec a(\vec v)$ [$\vec v$ being defined in \eq{lissajous}], which follows from the scattering matrix \cite{supplement}. Here, we restrict ourselves to a qualitative explanation and plot its component $a_1$ as a function of $\tvlr$ in \fig{fig1}{c}; note that $|a_2|\simeq |a_1|$. $\vec a(\vec v)$ vanishes unless $\mun\simeq\mulr$ and, hence, marks the charging lines. Its direction relates to the ratio $\gl/\gr$: since $a_1\simeq \pm a_2$, $\vec a$ is either parallel or anti-parallel to the $(1,1)$ direction in \fig{fig1}{c}, depending on whether $\tvl\gg\tvr$ (implying $\gl\gg\gr$) or $\tvl\ll\tvr$.

The third contribution in \eq{current}, $I^\text{rect}$, stems from a tiny ac modulation of the lead chemical potentials induced by the capacitive coupling between each gate and its adjacent lead. The amplitude of the resulting ac source-drain voltage is of the order $0.1\mu$V and, owing to the capacitive coupling, is phase-shifted by $-\pi/2$ with respect to the gate voltages.  This phase shift between the ac bias and the time dependent QD conductance gives rise to dynamic rectification.  $I^\text{rect}$ turns out essential for the quantitative agreement between the measured currents and the theoretical predictions of our model.  Interestingly, $I^\text{rect}\ne0$ requires a phase shift between the two modulated gate voltages. As such, dynamic rectification is very different in nature than the usually and hitherto discussed rectification caused by static asymmetries of the geometry and $I$-$V$ curve \cite{Giblin2013}.

The symmetries of these three contributions as a function of $\phi$ can be
revealed by formally reverting time \cite{supplement}. We find $2k$
symmetry points at which $\bar G$ obeys TRS, while $Q^\text{cycle}$ and
$I^\text{rect}$ are anti-symmetric under time reversal and, hence,
under phase reflection.

\Fig{fig1}{c} contains example Lissajous figures $\vkphi$ for $\phi=\pi/(k+1)$: a circle for $k=1$, a distorted figure eight for $k=2$ and a triple loop for $k=3$. For $k=1$ and $\phi$ slightly different from $\pi$, $\vec v_{k=1,\phi}(t)$ is an eccentric ellipse which corresponds to the pumping measurements already discussed in literature \cite{Ono2003,Jehl2013} [dashed line in \fig{fig1}{c}].  Whenever $\vkphi$ crosses a charging line, the charge of the QD changes by one electron. If this happens in a red region corresponding to $a_1,a_2>0$ with $\gl>\gr$ an electron will be exchanged preferably with the left lead.  In the blue region with $a_1,a_2<0$ and $\gl<\gr$, charge exchange with the right lead is preferred. We define $I>0$ for electrons flowing from the right to the left.  Then, crossing a blue area from above and a red area from below (see example for $k=1$) both contribute to $I<0$, each with half an electron charge per cycle. Our example for $k=3$ also results in $I<0$.  For $k=2$, the contributions of the outer loops to $Q^\text{cycle}$ have opposite sign and, thus, cancel each other to some extent, despite that the curves have a different symmetry axis than the resonance lines.  The same holds for the inner loops.  Therefore, we expect the pump current for $k=2$ to be generally smaller than for $k=1$.  For $k=3$, by contrast, symmetry-related parts of each curve again contribute to $Q^\text{cycle}$ with the same sign.  A generalization of these arguments leads to the expectation that for even values of $k$, the pump current should be larger than for odd values.  For larger $k$, however, there will be an increasing number of contributions with any sign and, thus, the situation becomes less clear.
\begin{figure*}[t]
\includegraphics[width=1\textwidth]{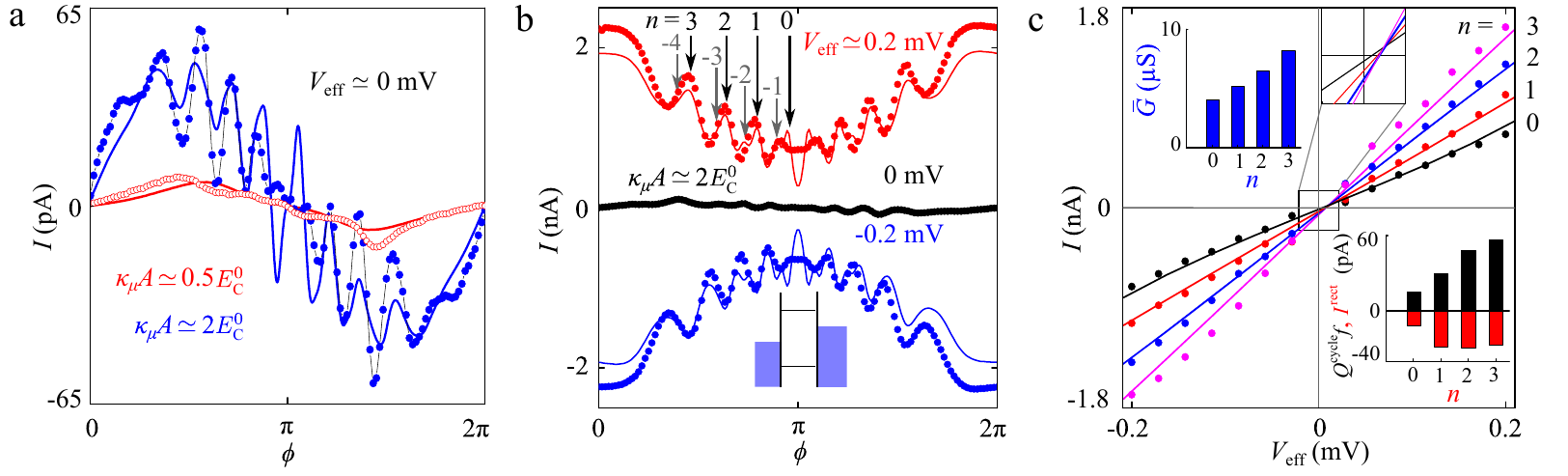}
\vspace{-5mm}
\caption{\label{fig3}
Symbols are measured, lines depict theory.
(a) $I(\phi)$ near zero bias for $\km A\simeq2E^0_C$, $V_\text{off}\simeq{-1}\,\mu$V (blue) and $\km A\simeq0.5E^0_C$, $V_\text{off}\simeq{0}\,\mu$V (red); $k=1$, $f=100\,$MHz.
(b) $I(\phi)$ for $\km A\simeq2E^0_C$, $V_\text{off}\simeq{9}\,\mu$V including curves at finite bias. Arrows indicate the phases at which the QD levels with indices $n$ start to reach the transport window ($\mul\le\mun\le\mur$ or $\mur\le\mun\le\mul$) during each modulation period. The relative shift between $n=0,-1$ and $n=1,-2$, etc.\ is caused by our working point not being centered between two QD chemical potentials [see \protect\fig{fig1}{a}].
(c) $I$-$V_\text{eff}$ characteristics for $\km A\simeq2E^0_C$, at current
peaks near black arrows in (b). The maxima slightly shift proportional to
$|V_\text{eff}|$. Upper right inset: magnification, theory lines intersect at
$V_\text{eff}\simeq{9}\,\mu$V. Upper left inset: average
conductance $\bar G$.  Lower inset: $I^\text{rect}$ (red, negative values) and pump current
$Q^\text{cycle}f$ (black, positive), both being independent of $V_\text{eff}$.
}
\end{figure*}%

In a realistic system, in addition, the symmetry of $\vec a(\vec v)$
is compromised because of point (iii) above: the separation of charging
lines, $\ec$, depends on $n$ and our working point (see \fig{fig1}{}) is
chosen to further decrease symmetry. As a result the contributions to
$Q^\text{cycle}$ from the outer loops for $k>1$ do not completely cancel
each other and we find a ratchet current for all $k$, specifically with
deviations from integer $Q^\text{cycle}$ for $k=3,5,7,\dots$ and from 
$Q^\text{cycle}=0$ for $k=2,4,6,\dots$. The mapping $\phi\rightarrow-\phi$
should survive the compromised symmetry discussed above and still result in
a reversal of the direction of $\vkphi$ and, therefore, also in a current
reversal.  Furthermore, the current direction should posses the $k$-fold
symmetry $\phi\to\phi+2\pi/k$. As a pump has two-fold symmetry, the case of
$k>1$ goes qualitatively beyond the scope of a pump.  For the $2k$ equally
spaced phases $\phi=0,\pi/k,\dots,(2k-1)\pi/k$, no charge is pumped at all,
$Q^\text{cycle}$ should change its sign and, thus, vanish. This symmetry is
apparent albeit not perfect in \fig{fig2}{} which compares the measured
ratchet currents for $k=2$ and $k=3$ as function of the modulation
amplitude and $\phi$  with model calculations. More values
of $k$ are available in \suppl. To improve modeling of our measured data we
included all experimentally known facts such as $\ec(n)$ and line
broadening into the numeric calculations (see \suppl). The best agreement
is then reached if we additionally assume a dc voltage of $V\sim\pm 5\mu$V.
This is likely the voltage offset of the used high-precision current
amplifier (input) which slowly drifts in time and is hard to control due to its dependence on the
ambient temperature.  Near the symmetry points of $Q^\text{cycle}=0$, the
symmetry of the overall current is markedly reduced by the dc-current $\bar
GV$ in \eq{current}, point (iv) above \cite{supplement}.

In \fig{fig3}{a} we present measured versus predicted $I(\phi)$-curves at $V_\text{eff}\simeq0$ for two different modulation amplitudes. For accuracy we have introduced the effective voltage $V_\text{eff}=V+V_\text{off}$ corrected by the voltage offset, $V_\text{off}$, caused by the current amplifier. At $\phi=\pi$, $\gl$ and $\gr$ oscillate in anti-phase and the QD states $\mun$ are static (for all $k$ and independent of $A$). The modulation of $\mun(t)$ grows with $|\phi-\pi|$ and with it the number of QD levels contributing to current, \ie fulfilling $\mun(t)=\mu_\text{L,R}$ twice during each modulation period. Consequently, between $0<\phi<\pi$ (equally for $\pi<\phi<2\pi$) $|I(\phi)|$ increases whenever another QD level starts to contribute to $Q^\text{cycle}$. For the smaller amplitude at most two levels with $n=-1,0$, for the larger amplitude at most eight levels with $-4\le n\le3$, reach $\mun=\mu_\text{L,R}$ during each period and, hence, contribute to $I$.  As $\phi$ approaches $0$ or $2\pi$ the gate modulation becomes symmetric restoring TRS, hence $I$ drops to zero \cite{supplement}. The differences between theory and experiments in \fig{fig3}{a} are not fully understood, but could be related to the contribution of excited states, which are not accounted for in our model.

As $V_\text{eff}$ is increased towards $|eV_\text{eff}|>{\ec^0}$ the dc contribution of the current $\bar G V_\text{eff}$ [first term of \eq{current}] rapidly gains weight yielding a transition to axial mirror symmetry reflecting the phase independent current direction determined by the sign of $V_\text{eff}$. This is clearly visible in \fig{fig3}{b} where we include data for the larger amplitude at finite $V_\text{eff}\simeq\pm0.2\,$mV.  The approximately quadratic current increase (decrease) away from $\phi=\pi$ (or equivalently from $\phi=0, 2\pi$) is related to the details of $\glr(t)$.

The symbols in \fig{fig3}{c} are the measured $I$-$V_\text{eff}$ characteristics at the current maxima near black arrows in \fig{fig3}{b}.  Here $|V_\text{eff}|<\ec^0/2e$ so that only a single QD state contributes to $I$ at any instance of time. In this regime we find approximately $I\propto{V_\text{eff}}$. Straight lines are the corresponding theory data which intersect as a common point at $V_\text{eff}\ne0$. The theory data are composed of three contributions shown as bar charts: $\bar G$ in blue, $Q^\mathrm{cycle} f$ in black and $I^\text{rect}$ in red. The fair agreement between theory and experiment confirms the theoretically assumed adiabaticity for our experiments.

In summary, we have realized a Lissajous rocking ratchet by modulating a single QD with two phase-locked voltage signals at commensurable frequencies. The directional motion is a consequence of  breaking the TRS which, as our analysis revealed, is restored only for certain phases between the signals. To achieve this, it is not sufficient to modulate the QD-lead couplings but additional modulation of the QD levels via capacitive cross-couplings is essential. While our experiments used a semiconducting QD, similar rocking ratchets could be realized in different systems, such as superconducting circuits, nano-electromechanical systems or molecular electronics. Lissajous ratchets encode the relative phase, frequency and amplitude information of two radio frequency signals into a time averaged dc signal, in spirit similar to a Lock-In amplifier. The ability to compare rf-signals on the chip promises a refined level of control desired for applications related to on-chip spectrum analyzing or quantum information processing.


We thank Klaus Pierz for supplying the wafer material.  We are grateful for financial support from the DFG via LU 819/4-1, SFB-631
and the Cluster of Excellence ``Nanosystems Initiative Munich (NIM)'' and
by the Spanish Ministry of Economy and Competitiveness via grant No.\
MAT2011-24331. S.L. acknowledges support via a Heisenberg fellowship of
the DFG.

\newpage
\appendix
\section{Additional experiments}

\subsection{Experimental setup}

\begin{figure}[ht]
\includegraphics[width=0.8\columnwidth]{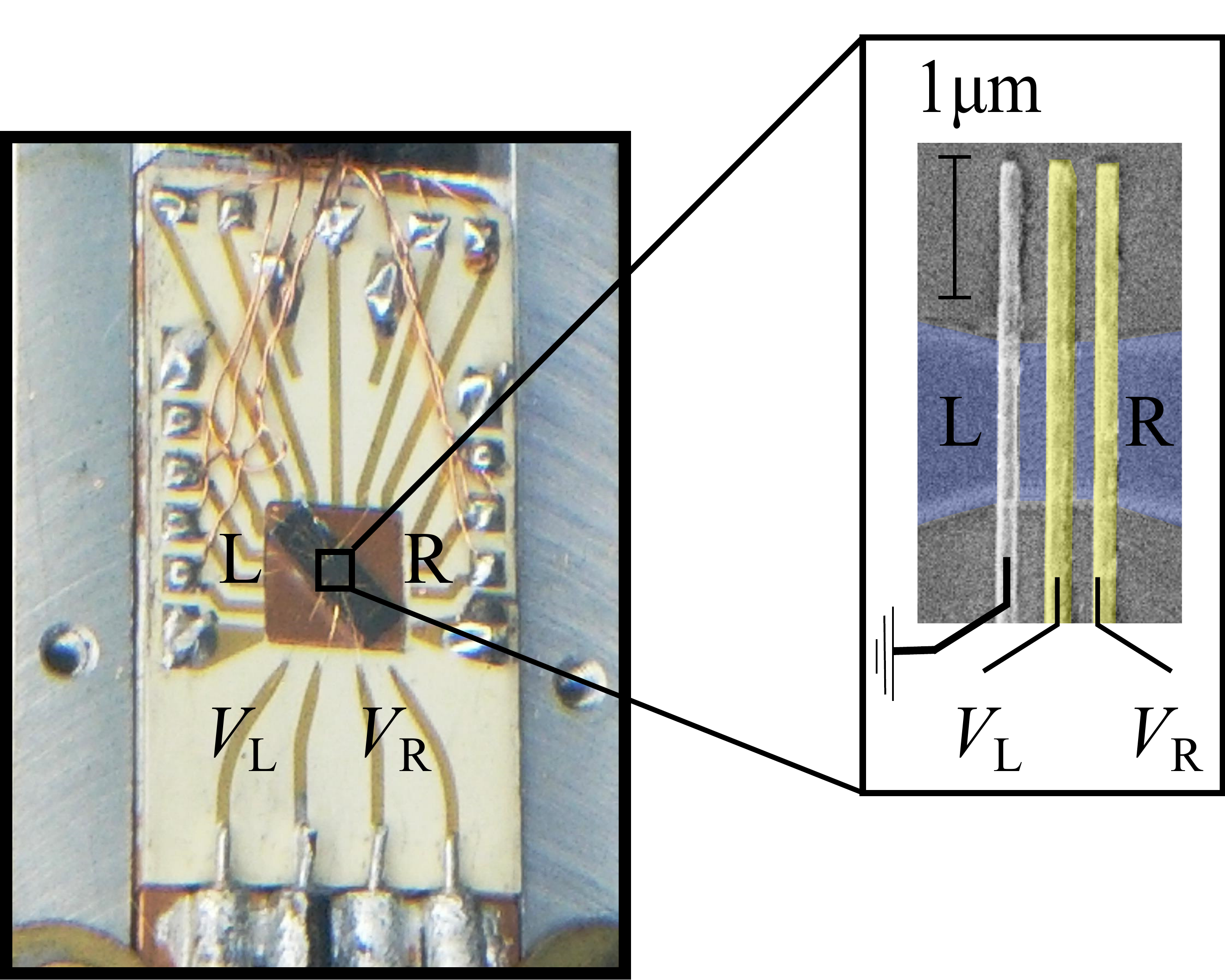} 
\caption{\label{fig:B1a} Radio frequency sample holder with sample and a
scanning electron microscope image of its surface (same image as in
\protect\fig{fig1}{a} of the main article).}
\end{figure}

\begin{figure}[h]
\includegraphics[width=0.8\columnwidth]{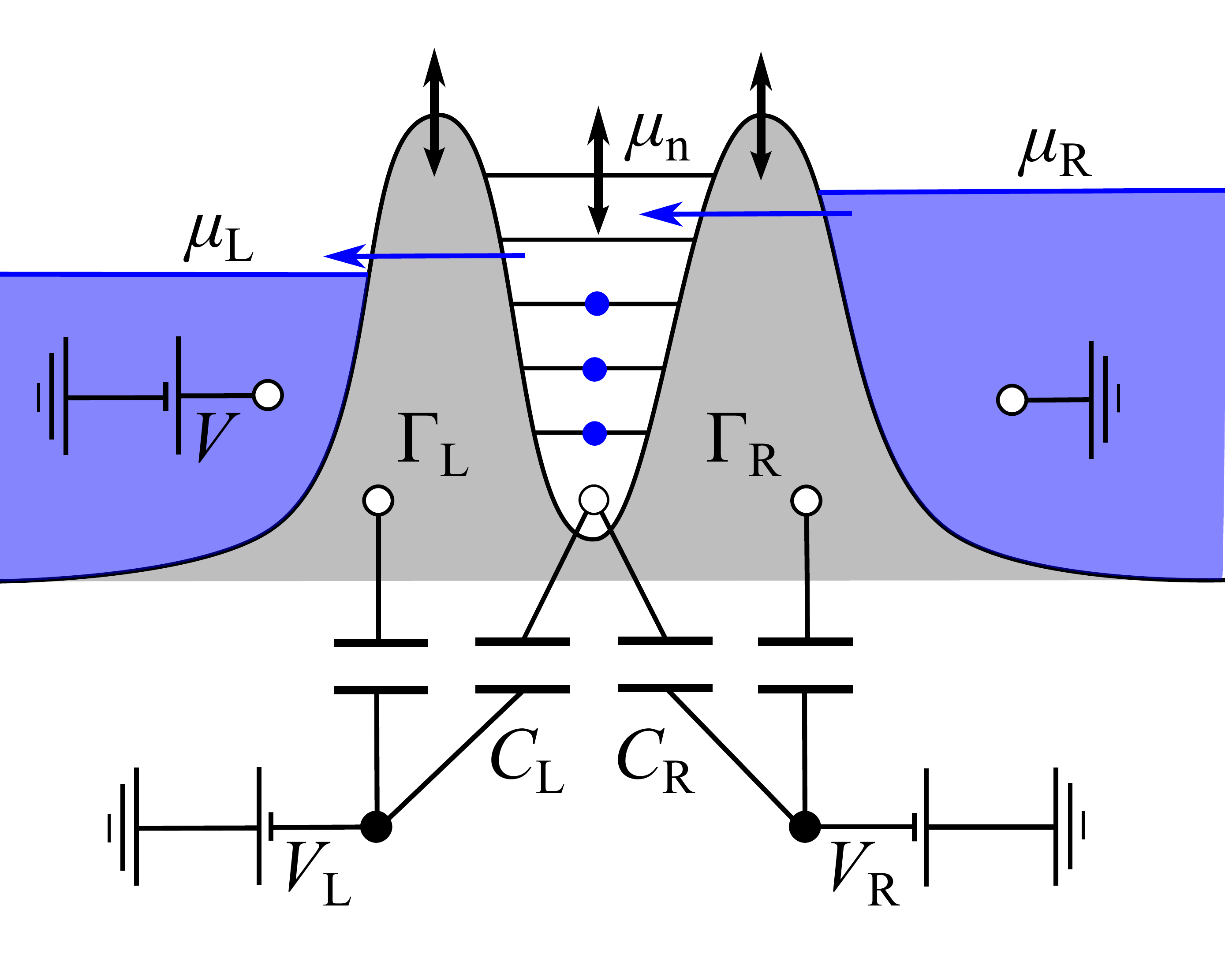} 
\caption{\label{fig:B1b}
Sketch of the QD circuit. Tunnel rates, $\glr$, and dot levels $\mu_n$ (horizontal lines) are capacitively controlled by the gate voltages, $\vl$ and $\vr$. The leads contain a degenerate electron system with chemical potentials, $\mulr$. The current is measured at the grounded right lead, while the voltage $V=(\mur-\mul)/e$ is applied on the left lead. For $V>0$ electrons tunnel via QD levels with $\mur\ge\mun\ge\mul$ from the right to the left lead (blue arrows).}
\end{figure}

The GaAs\,/\,AlGaAs wafer containing the QD sample was mounted on a radio frequency (rf) sample holder as shown in \fig{fig:B1a}{},
such that the wafer was in direct contact with a gold plated copper surface thermally connected to the mixing chamber of a dilution refrigerator at a base temperature of $\simeq50\,$mK. The control gates used to define the QD barriers (yellow gates in the detailed sample view in \fig{fig:B1a}{}, the gray gate is grounded) are connected to stainless steel rf coax cables. Source and drain contacts are connected via standard constantan wires and low pass filtered at room temperature. All cables are heat sinked at several points in the cryostat. A simplified sketch of the device circuit highlighting the capacitive control is presented in \fig{fig:B1b}{}.

\subsection{Calibration measurements}

\begin{figure}[ht]
\includegraphics[width=1\columnwidth]{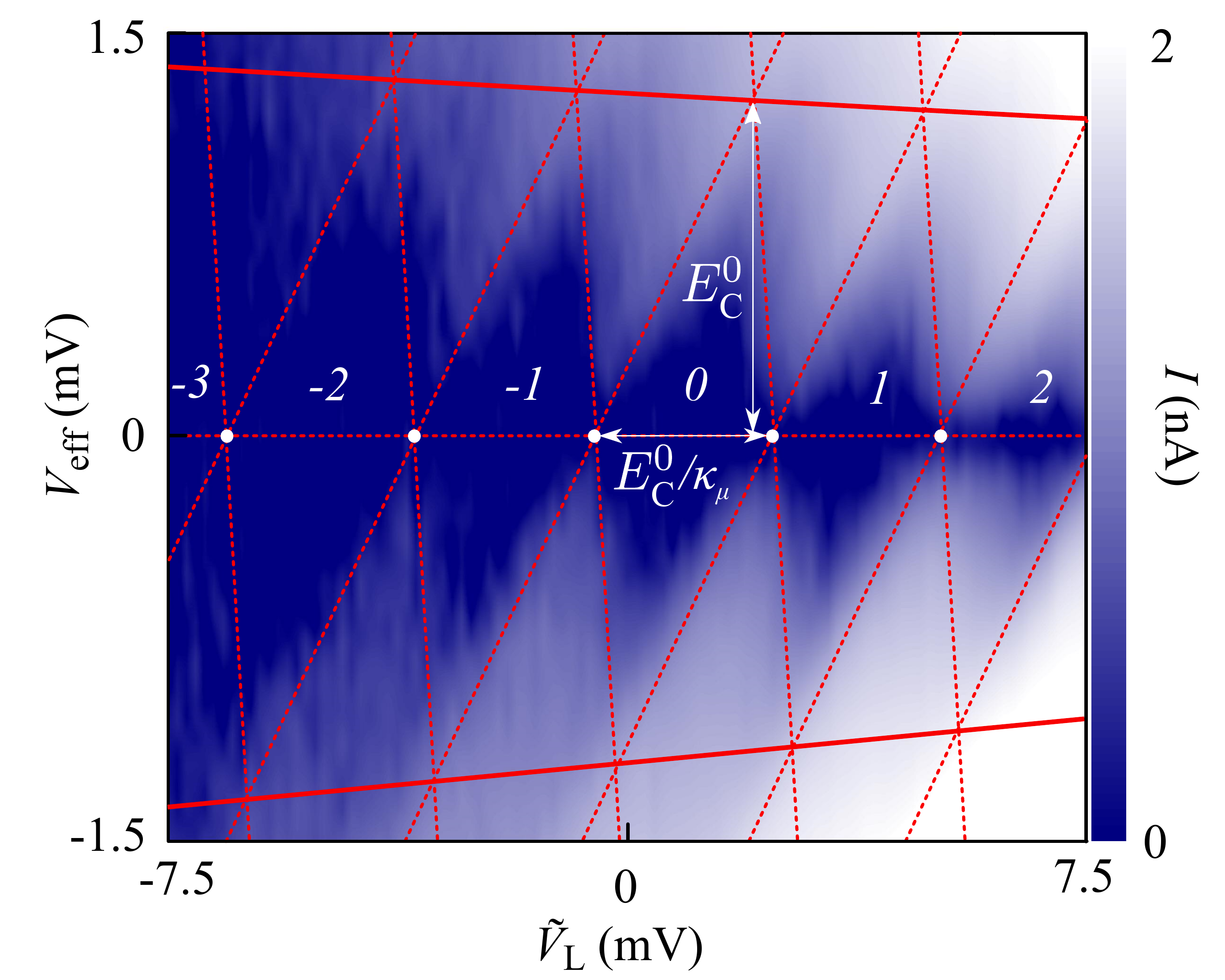}
\caption{\label{fig:B2} Calibration of charging energies and gate voltages to energy conversion. Gate voltages are not modulated. Plotted is the current through the QD measured as a function of gate voltage $\tvl=\vl+202\,$mV and bias $V_\text{eff}$ at constant $\tvr=\vr+91\,\text{mV}=-2.5\,$mV. The numbers indicate our index parameter $n$. White dots mark the corresponding current maxima positions in Fig.\ \ref{fig1}{a} of the main article. Red lines illustrate the Coulomb blockade diamonds.
}
\end{figure}

\begin{figure}[ht]
\includegraphics[width=1\columnwidth]{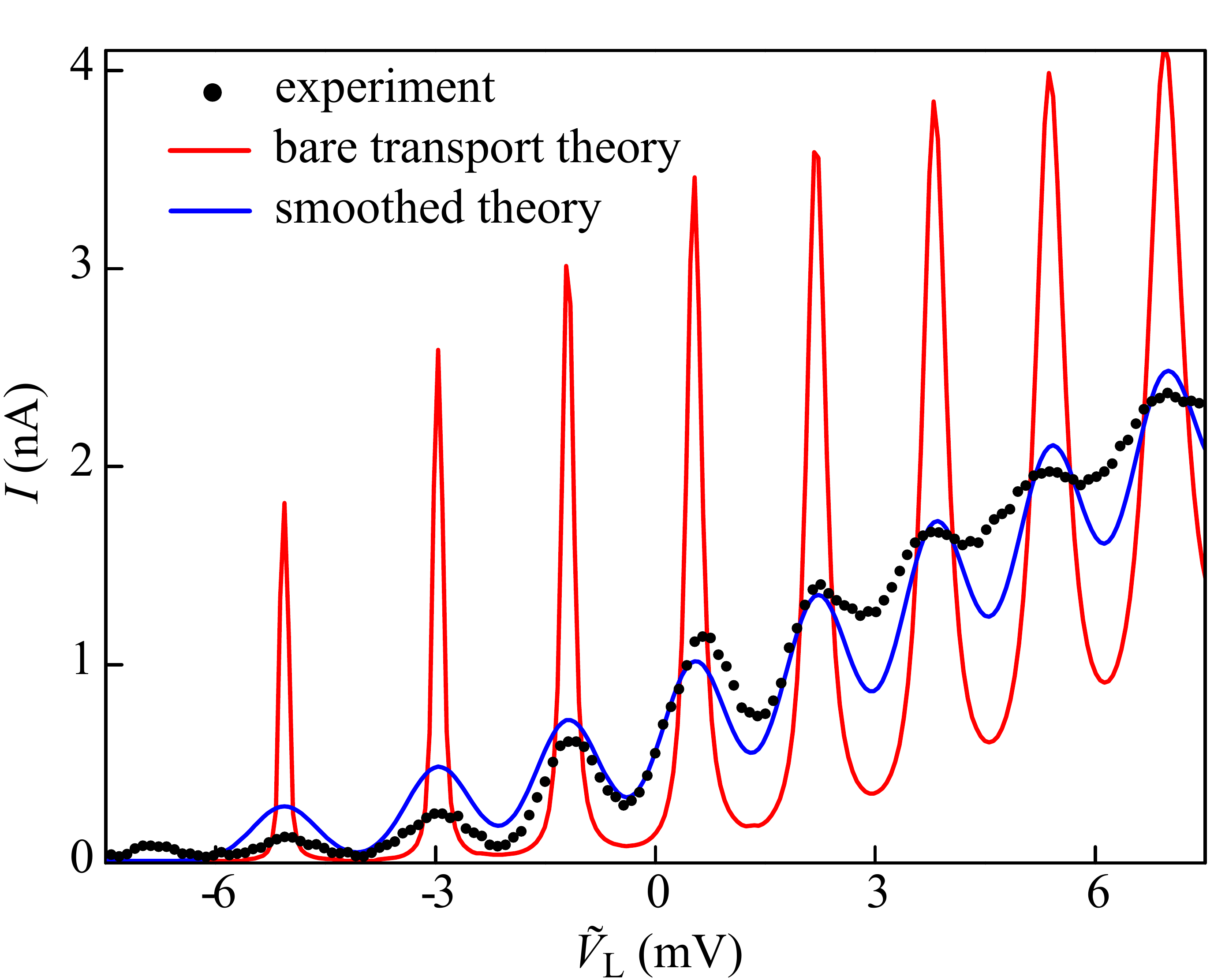}
\caption{\label{fig:B4}Calibration of tunnel couplings. Current through the QD for $V=100\,\mu$V, data from Fig.\ \ref{fig1} along its bisecting line where $\vl=\vr$ and $\gl\simeq\gr$ (black dots).  Also shown are two theory curves, all using $\kg=0.03e$, $\Gamma=0.13\,$meV in \eq{sens}, charging energies from Table \ref{tableI} and assuming $\gl=\gr$: $e\glr/h$ according to \eq{sens} current calculated with standard scattering theory with finite tunnel barriers as only broadening mechanism (red solid line), see \eq{app:IL} and below; calculated current as above but additionally convoluted with a Gaussian inhomogeneous distribution of constant width mimicing slow charge noise (blue line).}
\end{figure}

Our numerical calculations rely on accurate measurements of the QD
characteristics including the relevant energy scales, capacitive coupling
constants, tunnel rates and realistic line broadening. In the following we
present our calibration measurements. We begin by showing in \fig{fig:B2}{}
a typical single dot stability diagram depicting the current $I$ through
the QD as a function of the voltage applied to the left gate $\tvl$ and the
source-drain voltage $V$. Note that varying the right instead of the left
gate would result in a comparable plot. 
The white dots along the $V=0$ horizontal indicate the positions of current
maxima in Fig.~\ref{fig1} of the main article. As expected, they are
located at the crossings of the Coulomb diamonds at $V=0$.
From the dimensions of the Coulomb
diamonds (horizontal and vertical arrows) we estimate the charging energies
$\ec(n)=e^2/C_\text{QD}(n)$ and the capacitive coupling
$\km=eC_\text{L,R}(n)/C_\text{QD}(n)$, where $C_\text{QD}(n)$ is the
eigen capacitance of the QD [at the working point $C_\text{L}^0\simeq
C_\text{R}^0\simeq50$ aF and $C_\text{QD}^0\simeq130$ aF]. We find an approximate linear dependence
$\ec(n)=\ec^0+n\delta E$ with $\delta E=-0.04$ meV and at the working point
$\ec^0=1.27$ meV (solid lines in \fig{fig:B2}{}) and an approximately
constant leverage factor $\km=0.38e$.
Thus, the gate voltages shift the chemical potentials of the resonances as $\mun(\tvl,\tvr)=\mun^0+\epsilon_0(\tvl,\tvr)$  with $\mun^0=\mun(0,0)$ and the onsite energy
\begin{equation}
\label{app:mu(v)}
\epsilon_0(\tvl,\tvr) = -\km(\tvl+\tvr).
\end{equation}
The above relations can be reinterpreted as $\mun=n\ec^0-n(n+1)\delta E/2+\mun^0(n=0)$, where $\mun^0(n=0)=0.41\,$meV is the difference between the chemical potential of grounded leads and that of the lowest unoccupied dot-level at $\tvl=\tvr=0$. 
The influence of the gate voltages on the dot-lead tunnel rates can be
estimated from the dc current shown in \fig{fig1}{a} of the main text.  In
\fig{fig:B4}{},
we plot these data along the bisecting line $\tvl=\tvr$ of \fig{fig1}{a}. The Coulomb blockade oscillations display current maxima, but their smoothed average exhibits an exponential dependence on gate voltages below the working point $\tvl=\tvr=0$ devolving into a linear increase above. To capture this averaged behavior with only two free parameters, we assume the dependence as
\begin{equation}
\label{sens}
\Gamma_{\alpha}(\tva) = \frac{\kg \tva}{1-\exp({-\kg\tva/\Gamma})}\,,
\end{equation}
where $\alpha=\text L, \text R$, while $\Gamma\equiv\gl^0=\gr^0 = 0.13\,$meV
is the tunnel coupling at the working point and $\kg\simeq0.03e$ is the
slope for large $\tvl,\tvr$.  The red solid line in \fig{fig:B4}{} represents the prediction from
\eq{app:IL} below based on scattering theory  including the finite couplings
$\gl(V)=\gr(V)$ as only broadening mechanism [discussed below \eq{app:IL}].
Convoluting this curve with an additional Gaussian
broadening of constant width yields a better fit to the measured data (blue line).
The Gaussian expresses an inhomogeneous broadening caused by slow statistical
fluctuations of the QD levels (charge noise). In our case this
inhomogeneous broadening is the dominant broadening mechanism. The non-perfect fit is likely a
consequence of the increasing relevance of dissipation as the barriers open
up. Note that the values of  $\gl(V)=\gr(V)$ are predetermined
by the integral of the current peaks which is independent of additional
inhomogeneous broadening.  In Table \ref{tableI} we summarize the
functional dependences along the bisecting line in \fig{fig1}{a} of the
main article according to the equations above.
\begin{table}[t]
\begin{ruledtabular}
\caption{\label{tableI}
Experimentally determined parameters of our QD; from left to right: index
number $n$, chemical potentials at the working point $\tvl=\tvr=0$,
charging energies $\ec(n)=\mun-\mu_{n-1}$, tunnel couplings for the case of
symmetric coupling $\gl=\gr$ and $\tvl=\tvr$ such that $\mun=0$, i.e.,
$\glr=\ga(\mun^0/2\km)$, see Eqs.~\eqref{app:mu(v)} and \eqref{sens}.
}
\begin{tabular}{rccc}
$n$ 	& $\mun^0$ (meV) & $\ec(n)$ (meV) & $\glr$ (meV) at $\mun=0$\\
\hline
$-3$ & $-3.85$           & 1.39 & 0.39 \\ 
$-2$ & $-2.25$           & 1.35 & 0.59 \\ 
$-1$ & $-0.90$           & 1.31 & 0.82 \\
$0$  & $\phantom{-}0.41$ & 1.27 & 1.09 \\
$1$  & $\phantom{-}1.68$ & 1.23 & 1.40 \\
$2$  & $\phantom{-}2.91$ & 1.19 & 1.74 \\
$3$  & $\phantom{-}4.11$ & 1.15 & 2.13 \\
$4$  & $\phantom{-}5.31$ & 1.11 & 2.53 \\
\end{tabular}
\end{ruledtabular}

\end{table}

\subsection{Radio frequency calibration --- proof of adiabaticity}

\begin{figure}[ht]
\includegraphics[width=1\columnwidth]{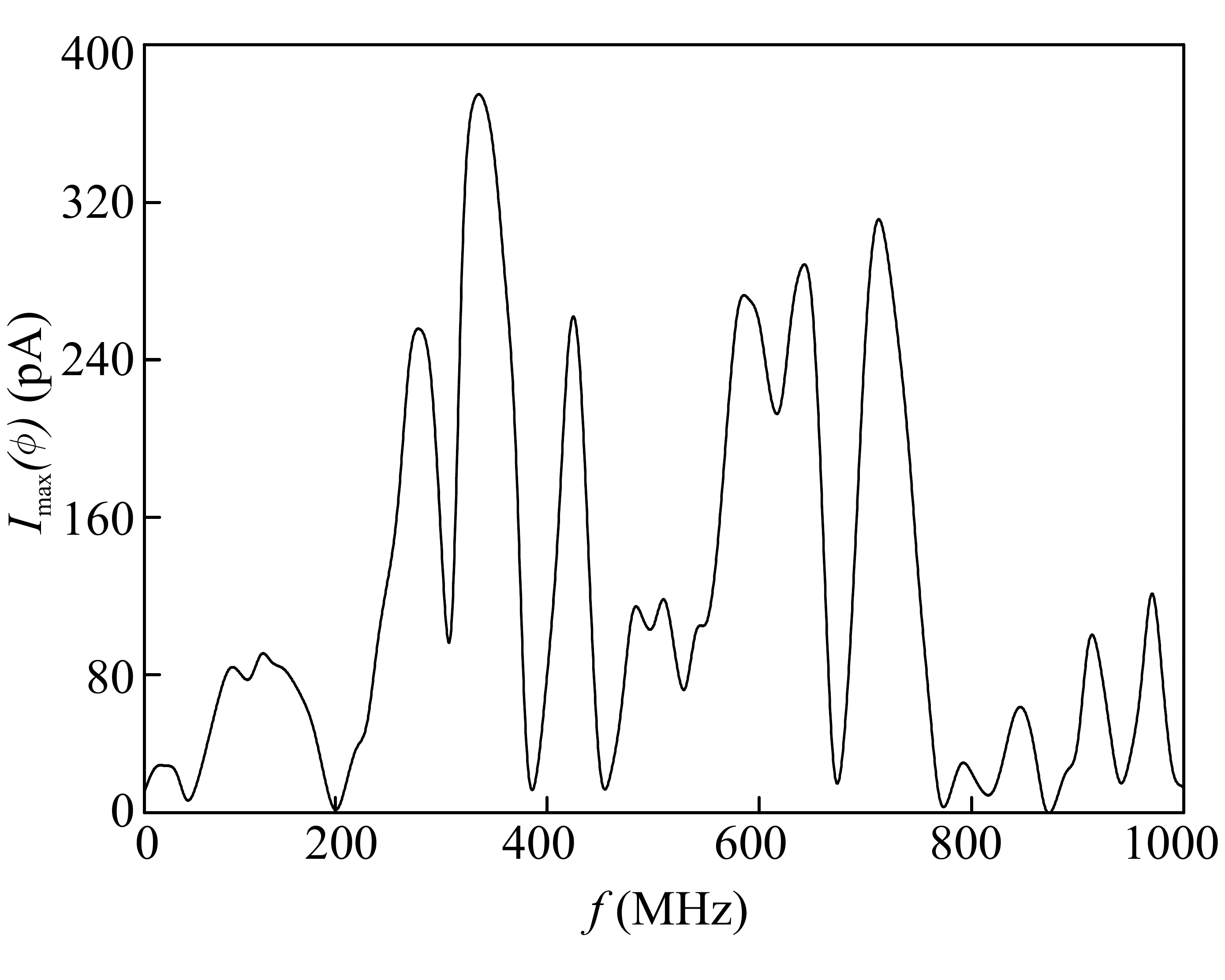} 
\caption{\label{fig:B5} Maximum current $I_\text{max}$ for phase difference $\phi$ from $0$ to $2\pi$ through the unbiased QD plotted as a function of modulation frequency at $k=1$ and modulation amplitude $A=3\ec^0$.
}
\end{figure}

\begin{figure}[ht]
\includegraphics[width=1\columnwidth]{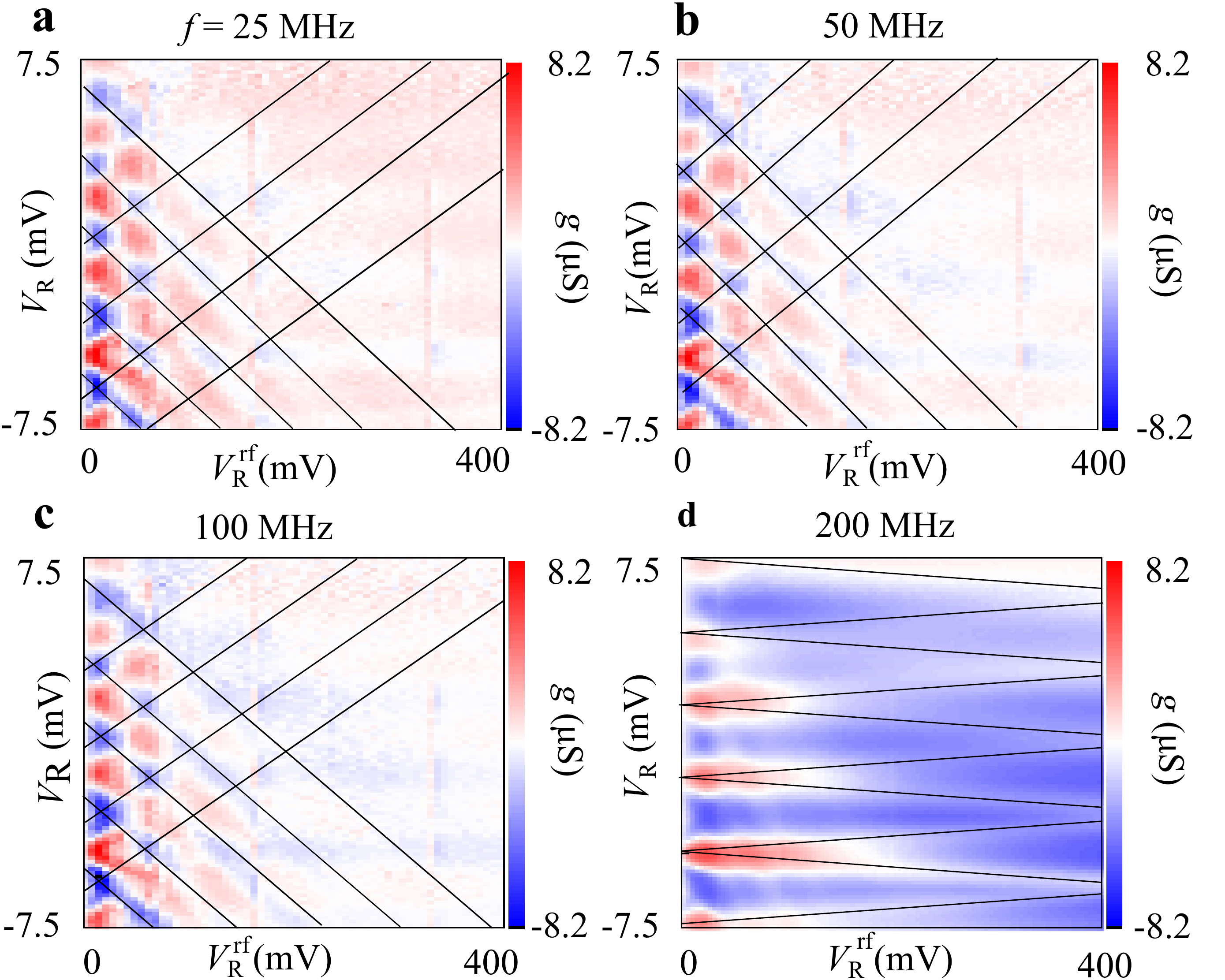} 
\caption{\label{fig:B6}
Radio frequency calibrations of the right hand side gate for $f=25, 50, 100, 200\,$MHz in panels a-d. Plotted is the differential conductance $g=dI/dV$ as a function of applied modulation voltage $\vr^\text{rf}$ and mean gate voltage $\vr$ (at constant $\vl=-90\,$ mV). The flocks of parallel black lines with mutually opposite slopes are guide for the eyes indicating current resonances which depart proportional to the increasing amplitude.
}
\end{figure}

\begin{figure}[ht]
\includegraphics[width=1\columnwidth]{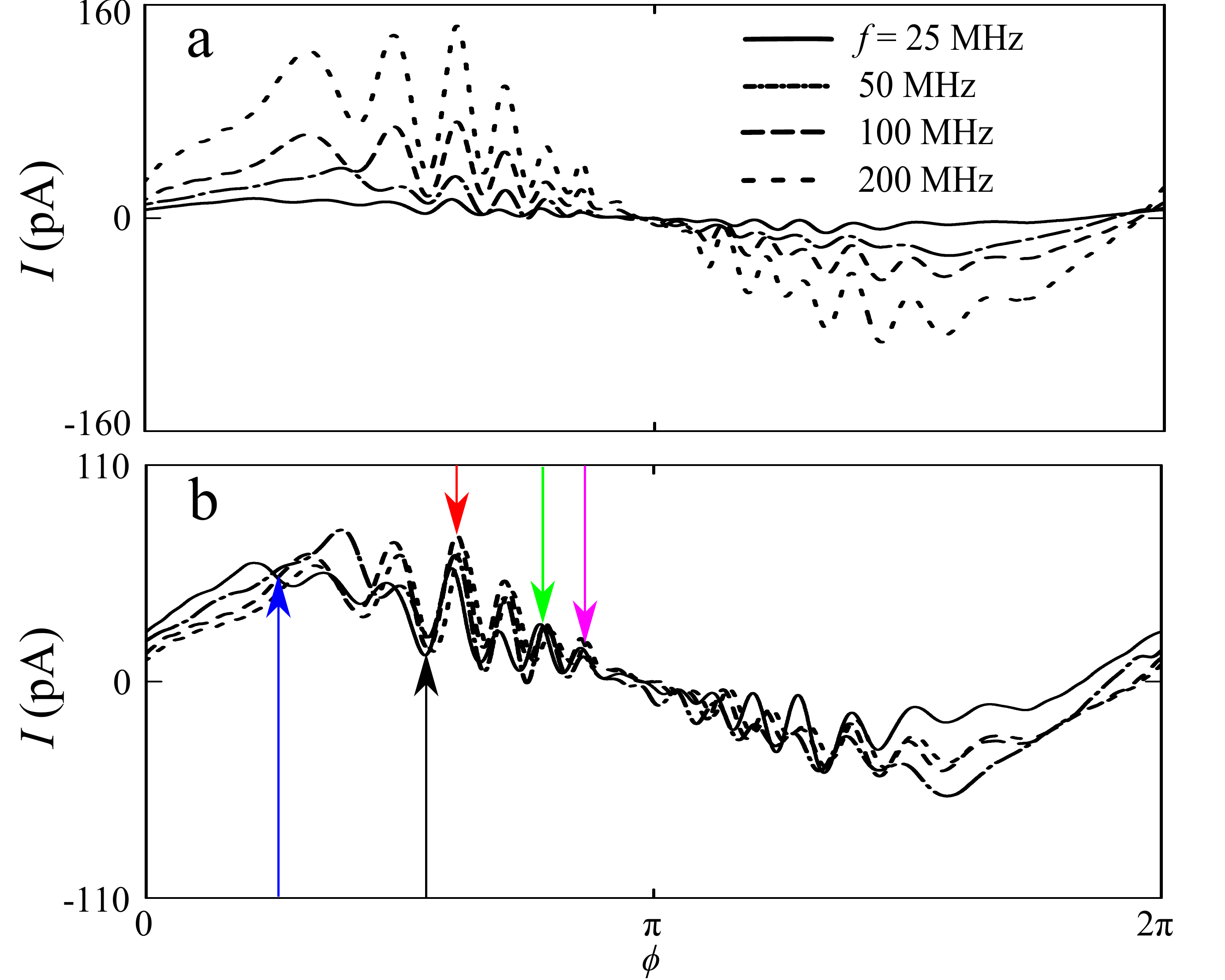} 
\caption{\label{fig:B7}
(a) Current $I$ as a function of phase difference $\phi$ for $k=1$, and $A=3\ec^0$. The current has been corrected for the frequency dependent transfer functions of the cables by applying the calibration factors $\alpha_{L,R}(f)$ listed in Table \ref{tableII} according to \eq{calibration}. (b) Same data as in panel a but scaled by $I(f)\to \frac{100\,\text{MHz}}{f}I(f)$.
}
\end{figure}

\begin{figure}[ht]
\includegraphics[width=1\columnwidth]{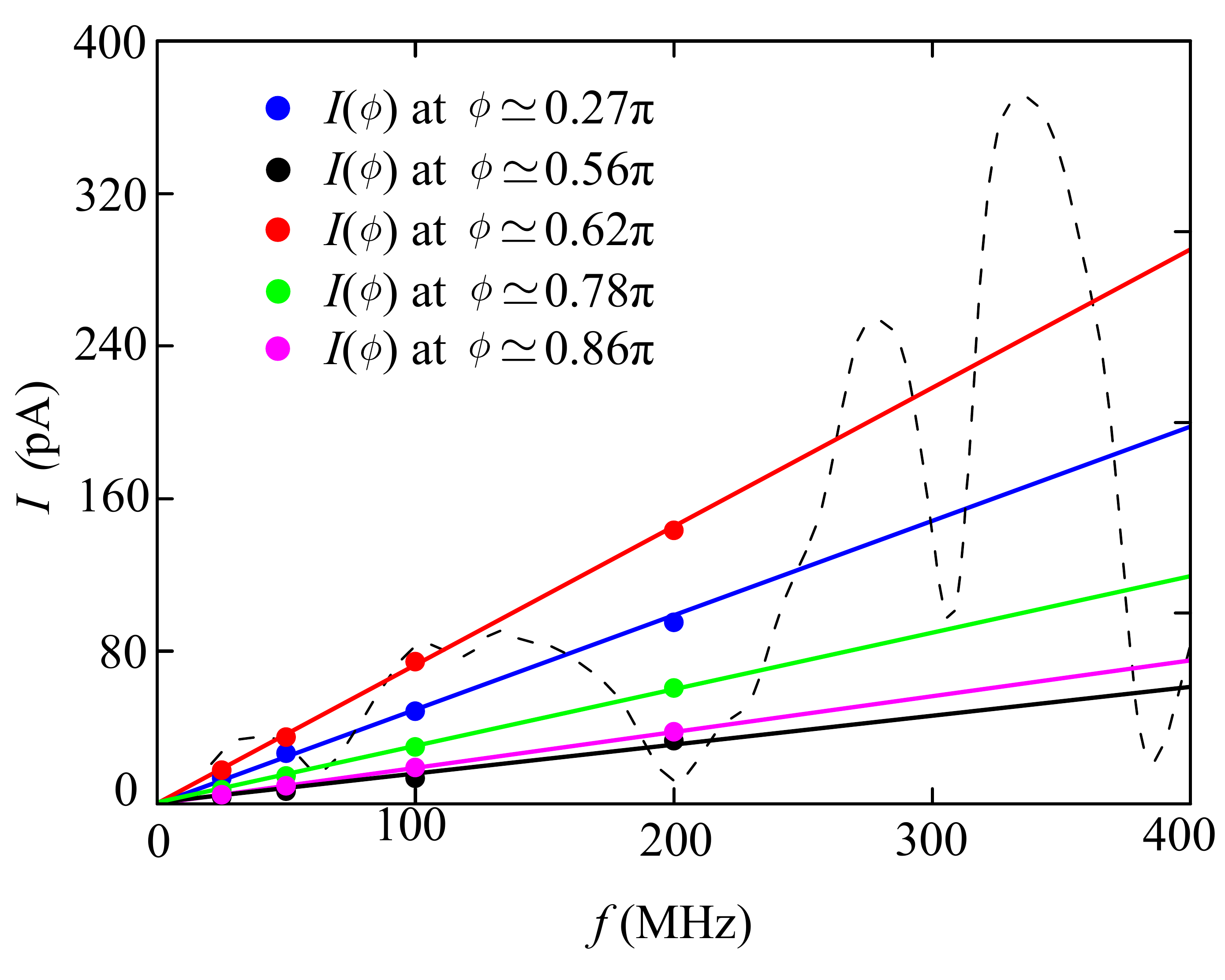} 
\caption{\label{fig:B8}
Frequency dependence of current before and after calibration of the modulation amplitude. The dashed line presents uncalibrated data identical to those in Fig.\  \ref{fig:B5}. Colored symbols indicate current values of specific maxima and minima (constant phase) marked in Fig.\ \ref{fig:B7}(a) by arrows; same as in the inset of Fig.\ \ref{fig1}(b) of the main paper. The straight lines are a guide for the eyes and verify the expected adiabatic regime.
}
\end{figure}

Next, we discuss the calibration of the amplitude of the radio frequency gate voltage modulation. Figure \ref{fig:B5}{}
demonstrates an apparent problem of our non-perfect rf setup, namely strong cable resonances, \ie standing waves caused by reflections of the rf-signal at the sample holder and meters away at the rf generator. In the adiabatic regime we expect to find a pump current $|I|\propto f$, if $V=0$. Instead we observe strong current oscillations which indicate an oscillating modulation amplitude even though the rf signal strength is fixed, the main reason being cable resonances. For our ratchet experiments we restricted ourselves to specific modulation frequencies, namely $f=25, 50, 100, 150, 200$\,MHz. In \fig{fig:B6}{} 
we demonstrate exemplarily for one of the gates the calibration of the modulation amplitude at these frequencies. Plotted is the differential conductance of the QD as function of $\vr$ and its modulation amplitude $\vr^\text{rf}$ (where $\vr(t)=\vr+\vr^\text{rf}\cos 2\pi f t$) while the left gate voltage is kept constant.  Each Coulomb resonance splits in two resonances located at the turn around points of $\vr(t)$. These splittings are proportional to $\vr^\text{rf}$ and are indicated by black lines in \fig{fig:B6}{}. Their slopes $\alpha_\text R(f)=\vr/\vr^\text{rf}$ can be used to calibrate the actual rf modulation amplitude according to:
\begin{equation}\label{calibration}
A_\text{L,R}=\km\alpha_\text{L,R}(f)\vlr^\text{rf}{/e}
\end{equation}
The result of the calibration (of both gates) is summarized in Table \ref{tableII}.
\begin{table}
\caption{\label{tableII}Radio frequency calibration factors $\alpha_{L,R}$ at various driving frequencies.}
\begin{ruledtabular}
\begin{tabular}{ccc}
$f$ in MHz &  $1/\alpha_L$ & $1/\alpha_R$\\
\hline 
25 & 25.2 & 24.9 \\
50 & 23.2 & 23.1 \\
100 & 27.5 & 27.2 \\
150 & 26.3  & 26.1 \\
200 & 182.5 & 178.6 \\
\end{tabular}
\end{ruledtabular}
\end{table}
To account for the frequency dependent calibration in our measurements we correct the amplitudes at each frequency according to Table \ref{tableII} and \eq{calibration}. In \fig{fig:B7}{a}
we present $I(\phi)$ at four frequencies for $k=1$ and identical modulation amplitudes $A=3\ec^0$ after calibration. The same data scaled by $I(f)\to \frac{100\,\text{MHz}}{f}I(f)$ to the current expected at $f=100\,$MHz are approximately frequency independent [\fig{fig:B7}{b}]. It demonstrates not only the validity of our procedure to calibrate the rf amplitudes but also corroborates our assumption of adiabatic charge transport. Small deviations, especially those near $\phi=0$ are probably related to a local disorder potential which compromises the spacial symmetry. The importance of a correct calibration becomes evident in \fig{fig:B8}{}
plotting $I_\text{max}$ in direct comparison to the uncalibrated current taken from \fig{fig:B5}{}. Straight lines are guide for the eyes indicating the adiabatic transport regime.

\section{Theoretical approach}
\label{app:formalism}

We derive a scattering formalism for the dc current through a periodically
driven conductor in the low-frequency limit, which generalizes the
scattering approach for parametric pumping \cite{Pretre1996a, Brouwer1998a}
to the presence of a finite dc voltage and a tiny ac bias.  The derivation
starts from Floquet scattering theory \cite{Wagner1999a, Moskalets2002a,
Camalet2003a, Kohler2005a} which we expand for a small driving frequency
$f=\Omega/2\pi$.  Despite that we will finally apply the approach to a
two-terminal device, we formulate the general theory for an arbitrary
number of leads.

A cornerstone of Floquet scattering theory is an expression for the
probability $T_{\alpha\beta}(t,\epsilon)$ that an electron with initial
energy $\epsilon$ from lead $\beta$ is scattered to lead $\alpha$ at
time $t$.  Then the time-dependent net current in lead $\alpha$ can be
expressed as \cite{Camalet2003a, Kohler2005a}
\begin{equation}
\label{app:ILt}
I_\alpha = \frac{e}{h} \int d\epsilon\Big[ f(\epsilon-\tilde\mu_\alpha(t))
- \sum_{\beta} T_{\alpha\beta}(t,\epsilon) f(\epsilon-\tilde\mu_\beta(t))
\Big] ,
\end{equation}
where $h=2\pi\hbar$ denotes Planck's constant while $f(x)=[\exp(
x/k_BT)+1]^{-1}$ is the Fermi function for thermal energy $k_BT$.  
In our experiment, the modulation of the gate voltages induces a tiny
ac bias caused by a tiny capacitance between the modulated gates and the
electrical leads of the QD. To capture this, we allow for periodically
time-dependent chemical potentials $\tilde\mu_\alpha(t) = \mu_\alpha
+ew_\alpha(t)$. Our goal is to derive a convenient expression for the
time-average of $I_\alpha(t)$ in the adiabatic limit.

We start by Taylor expansion of the Fermi functions in the
ac shift $w_\alpha(t)$. It yields the time-dependent current
$G_{\alpha\beta}(t)w_\beta(t)$ with the time-periodic conductance matrix
\begin{equation}
\label{app:Gt}
G_{\alpha\beta}(t) = G_{\alpha\beta}(\vec v(t))
= \frac{e^2}{h}
\left[\delta_{\alpha\beta}-T_{\alpha\beta}(t,\bar\mu)\right].
\end{equation}
Despite that the ac bias $w_\beta(t)$ vanishes on average, its impact
becomes rectified and contributes
\begin{equation}
\label{app:Irect}
I_\alpha^\text{rect}
= \int_0^T \frac{dt}{T}\,G_{\alpha\beta}(\vec v(t)) w_\alpha(t)
\end{equation}
to the dc measured current, where $T$ is the modulation period.

Having considered the ac bias, we now come to the effects of a dc voltage
and the modulation of tunnel barriers and the QD levels. We employ
ideas put forward in Refs.~\cite{Moskalets2002a, EntinWohlman2002a,
Kashcheyevs2004b}. For the now constant chemical potentials, the
time-average of the transmission $T_{\alpha\beta}(t,\epsilon)$ provides
$T_{\alpha\beta}^{(k)}(\epsilon)$, which is the probability that an
electron with initial energy $\epsilon$ coming from lead $\beta$ is scattered to
lead $\alpha$ after the absorption of $k$ energy quanta, where negative $k$
corresponds to emission.  Thus, the final energy of the electron is
$\epsilon+k\hbar\Omega$.  Then the net average current flowing from lead
$\beta$ to the conductor assumes the intuitive form \cite{Wagner1999a,
Moskalets2002a, Camalet2003a, Kohler2005a}
\begin{equation}
\label{app:IL}
I_\alpha = \frac{e}{h} \int d\epsilon\Big[ f(\epsilon-\mu_\alpha)
- \sum_{\beta,k} T_{\alpha\beta}^{(k)}(\epsilon) f(\epsilon-\mu_\beta)
\Big] .
\end{equation}
This expression can be derived from a time-dependent wire-lead model which
relates the transmission to the scattering matrix and the Green's function
of the driven conductor \cite{Wingreen1993a, Moskalets2002a, Kohler2005a}.
From the unitarity of the scattering matrix follows charge conservation,
which means that an electron with initial energy $\epsilon$ will be
scattered with probability unity to any sideband $k$ in any lead $\alpha$,
so that the sum over all these channels must fulfill the relation
\begin{equation}
\label{app:kirchhoff1}
\sum_{\alpha,k} T_{\alpha\beta}^{(k)}(\epsilon) = 1 .
\end{equation}
With the same arguments but summing over all initial states that end up
with energy $\epsilon$ in lead $\alpha$ we obtain a further sum rule:
\begin{equation}
\label{app:kirchhoff2}
\sum_{\beta,k} T_{\alpha\beta}^{(k)}(\epsilon-k\hbar\Omega) = 1 .
\end{equation}

\subsection{Adiabatic limit}

We have performed experiments in the adiabatic limit corresponding to an
expansion of $I_\alpha$ to first order in $\Omega$. For the Taylor
expansion we insert the left-hand side of Eq.~\eqref{app:kirchhoff2} as
prefactor of $f(\epsilon-\mu_\alpha)$ into the current formula
\eqref{app:IL} and obtain
\begin{equation}
\begin{split}
\label{app:IL2}
I_\alpha
= \sum_{\beta,k} \frac{e}{h} \int d\epsilon \Big\{ &
  T_{\alpha\beta}^{(k)}(\epsilon) [ f(\epsilon-\mu_\alpha)-f(\epsilon-\mu_\beta)]
\\ &
  -k\hbar\Omega \frac{\partial T_{\alpha\beta}^{(k)}(\epsilon)}{\partial\epsilon}
  f(\epsilon-\mu_\alpha) \Big\} .
\end{split}
\end{equation}
Notice that in contrast to Ref.~\cite{Moskalets2002a}, we have expanded in
$\Omega$ for the transmission rather than for the Fermi
function, so that the result holds whenever the transmission as a
function of $\epsilon$ is smooth on the scale $\hbar\Omega$.  In
particular, it is valid also for low temperature, i.e.\ beyond the
high-temperature limit of $k_BT\gg\hbar\Omega$.

Next we make use of the time-periodicity of the driving which has the
consequence that the propagator $U(t,t-\tau)$
depends not only on the difference $\tau$, but explicitly on both the
initial time $t-\tau$ and the final time $t$.  After
Fourier transformation with respect to $\tau$, one obtains both the
propagator and the scattering matrix for an electron with initial energy
$\epsilon$ \cite{Kohler2005a}.  The result is periodic in time and can be
written as
\begin{equation}
S(t,\epsilon) = S(t+T,\epsilon) = \sum_k e^{-ik\Omega t} S^{(k)}(\epsilon) .
\end{equation}
Then from the usual relation between transmission amplitudes and
probabilities, $T_{\alpha\beta}(t,\epsilon) = |S_{\alpha\beta}(t,\epsilon)|^2$,
and taking the time-average of the current follows
$T_{\alpha\beta}^{(k)}(\epsilon) = |S^{(k)}_{\alpha\beta}(\epsilon)|^2$.
[Notice that generally the transmission to the $k$th sideband,
$T_{\alpha\beta}^{(k)}(\epsilon)$, is different from the $k$th Fourier
coefficient of $T_{\alpha\beta}(t,\epsilon)$].  Next we employ Parseval's
theorem to write the $k$-summation in Eq.~\eqref{app:IL2} as
time-integration over one driving period.
Moreover, it is convenient to define voltages $V_\alpha$ as deviations from
an average chemical potential, $\mu_\alpha = \bar\mu+eV_\alpha$, and evaluate
the Fermi functions for zero temperature and small voltage.  Then we end up with
the time-averaged current
\begin{equation}
\label{app:IL3}
I_\alpha = \sum_\beta \bar G_{\alpha\beta} V_\beta
+I_\alpha^\text{rect} + \frac{Q_\alpha^\mathrm{cycle}}{T},
\end{equation}
with the time averaged conductance
\begin{equation}
\label{app:Gbar}
\bar G_{\alpha\beta} = \int_0^T \frac{dt}{T}\,G_{\alpha\beta}(\vec v(t))
\end{equation}
and the current resulting from the rectified ac bias given in
\eq{app:Irect}.
The last term in \eq{app:IL3} reflects the charge
parametrically pumped through contact $\alpha$ during one driving period,
\begin{equation}
\label{app:Qt}
Q_\alpha^\mathrm{cycle}
= \frac{1}{2\pi} \mathop{\mathrm{Im}}
  \int_0^T dt \Big(S(t,\mu_\alpha) \frac{\partial}{\partial t}
  S^\dagger(t,\mu_\alpha)\Big)_{\alpha\alpha} .
\end{equation}

The shape of the driving can be expressed by a closed curve $\mathcal{C}$ in
parameter space, $\vec v(t) = \vec v(t+T)$.  Then, if $\mathcal{C}$ is traversed
adiabatically slowly, the scattering matrix depends merely parametrically
on time, \ie $S(t,\epsilon) = S(\vec v(t),\epsilon)$, where the latter is
the static result for the instantaneous value of $\vec v$ at time $t$.
This allows one to transform the time integral in Eq.~\eqref{app:Qt} to a
closed line integral to obtain
\begin{equation}
\label{app:Qx}
Q_\alpha^\mathrm{cycle}
= \frac{1}{2\pi} \mathop{\mathrm{Im}}
  \oint_\mathcal{C} d\vec v \cdot \big(S(\vec v,\mu_\alpha) \mathop{\mathrm{grad}}
  S^\dagger(\vec v,\mu_\alpha)\big)_{\alpha\alpha} .
\end{equation}
Applying Stokes' theorem, one can transfer this expression into a surface
integral to obtain Brouwer's formula \cite{Brouwer1998a}.  For the
numerical evaluation in the parameter range considered here, however, the
present form is more appropriate.

An important implication of Eq.~\eqref{app:IL3} is the separation of the
average current into a dc contribution stemming from the average
conductance and an adiabatically pumped charge.  The main difference
between these two quantities is their behavior under time inversion of the
closed curve $\mathcal{C}$: While the average conductance is invariant,
the pumped charge acquires a minus sign.  Below, we will explore this
symmetry property for the Lissajous curves applied in the experiment.

\subsection{Independent channel approximation}

In our experiment, the conductor is formed by a QD connecting
two leads, $\alpha=L,R$.  It is driven with an amplitude $A$ ranging from
zero to the rather large value $5\ec^0$, so that the onsite energy of the
QD may change by several charging energies---we will consider up
to 10 excess electrons.  Moreover, the influence of the gate voltage applied to
the tunnel barriers turns out to be crucial for the observed pumping, so
that a master equation approach based on lowest-order perturbation theory
in the dot-lead coupling is not appropriate.  As a consequence, already for
the static transport problem, a full treatment that includes all possible
spin and correlation effects is practically impossible.  We instead assume
that each charge state of the quantum dot contributes as independent
transport channel to the current.  This requires that the separation of the
resonances given by the charging energy is larger than their
homogeneous widths such that the conductance peaks are well separated.
Moreover, the driving must be slow enough that the quantum dots can be
assumed to be always relaxed to its many-particle ground state.
\begin{figure*}[ht!]
\centerline{\includegraphics[scale=1]{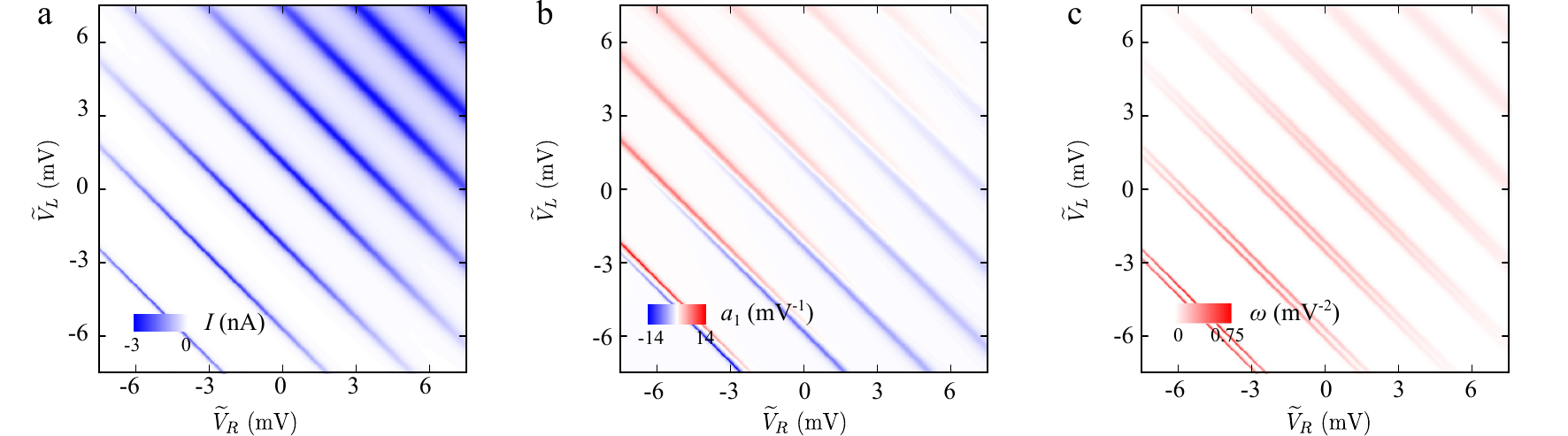}}
\caption{
(a)
Current in the absence of the modulation for bias voltage $V=200$mV.
(b)
Component $a_1(\vec v)$ of the vector field \eqref{app:vectorfield}
appearing in the symmetrized line integral representation of the charge
pumped per cycle.  Positive values (blues stripes, marked with $+$) emerge
when a QD level equals the chemical potential of the left lead and
correspond to electron pumping from the left lead to the QD.  Accordingly,
the curvature is negative (red, $-$) when a QD level agrees with the
chemical potential of the right lead, while the mismatch stems from the
voltage bias. In the relevant regime, the second component, $a_2(\vec v)$,
looks very similar.
(c)
Corresponding curvature $\omega = (\partial a_1/\partial x_2-\partial a_2/\partial v_1)$ that appears in the symmetrized Brouwer's formula.  The lines are splitted due to the bias.
}
\label{fig:I,ax}
\end{figure*}
Our QD model is parametrized by the energies $\mun$ necessary for adding
the next electron to the QD.  Thus, the position of the
corresponding conductance peak is directly related to $\mun$.  Assuming that
all these peaks are well separated, we can treat the resonances as
independent levels that can be occupied by a spinless electron.  The
corresponding scattering matrices read
\begin{equation}
\label{app:Sn}
S_n(\epsilon) = \mathbf{1} -\frac{i}{\epsilon-\mu_n+i(\gl+\gr)/2}
\begin{pmatrix} \gl & \sqrt{\gl\gr}\\
\sqrt{\gl\gr} & \gr \end{pmatrix} ,
\end{equation}
where $\gl$ and $\gr$ are the dot-lead tunnel rates.

We drive our QD by applying time-dependent gate voltages to the tunnel
barriers.  Thus these gate voltages form our parameter space and are the
components of the vector $\vkphi\equiv [\tvl(t),\tvr(t)]$, see Eq.~(1) of
the main article.  They entail a time-dependence on the
onsite energy $\epsilon_0$ and thus on $\mu_n$ as well as on the dot-lead
tunnel rates $\gl$ and $\gr$.  For the onsite energy, we assume
the linear relation \eqref{app:mu(v)}.
The height of each tunnel barrier is also shifted approximately linearly by
the respective gate voltage.
For ideal tunnel barriers according to the WKB formula, one would
expect that the dot-lead tunnel rates depend exponentially on the gate
voltages.  Our dc measurements, however, indicate that this overestimates
the rates for large positive $\vlr$.  Therefore, we assume for
$\Gamma_{\alpha}(\vec v)$ the dependence given by \eq{sens} with
$\Gamma\equiv\gl^0=\gr^0 = 0.07\,$meV and $\kg\simeq0.03$.
In order to evaluate the current formula \eqref{app:IL3}, we insert the
scattering matrix $S_n$ into Eqs.~\eqref{app:Gbar} and \eqref{app:Qx} and
subsequently sum the contribution of the resonances $n$.

Even though this approach already allows us to obtain the dc current (all
our numerical results are computed in this way), it is instructive to
investigate the contribution of a single resonance to both the average
conductance and the charge per cycle.  For this purpose, we assume that at
a resonance, the dot-lead rates are weakly time-dependent with average
values $\bar\Gamma_{L,R}$, while $\mu_n(t)\simeq (t-t_0)\dot\mu_n$.  Then
the straightforward evaluation of Eq.~\eqref{app:Gbar} with the
off-diagonal matrix element of $S_n$ yields
\begin{equation}
\bar G^n \simeq \frac{e^2}{h} \frac{\bargl\bargr}{\bargl+\bargr}
\frac{\Omega}{|\dot\mu_n|}
\end{equation}
Notice that $\dot\mu_n$ is proportional to $\Omega$ and, thus, $\bar G^n$ is
frequency independent.

\begin{figure*}[ht!]
\centerline{\includegraphics[width=1\textwidth]{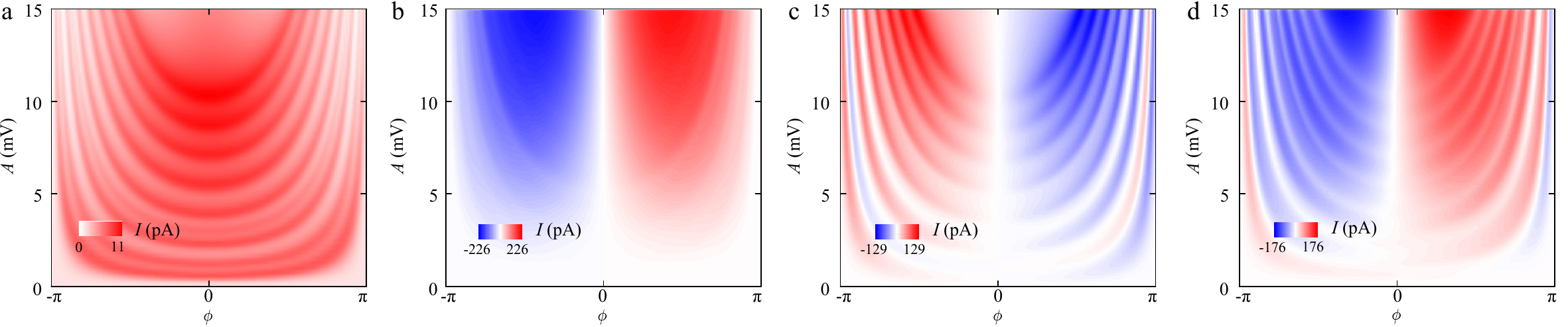}}
\caption{
Theoretical prediction of the current for $k=1$ shown in \protect\fig{fig1}{b}
of the main article, $f_\text L=f_\text R=200\,$MHz.
The dc tunnel current $\bar G V$ (a), the pumped current $Q^\text{cycle}f$ (b), 
and the ac current (c) sum up to the total current plotted in (d).
}
\label{fig:k1}
\end{figure*}
\begin{figure*}[ht!]
\centerline{\includegraphics[width=1\textwidth]{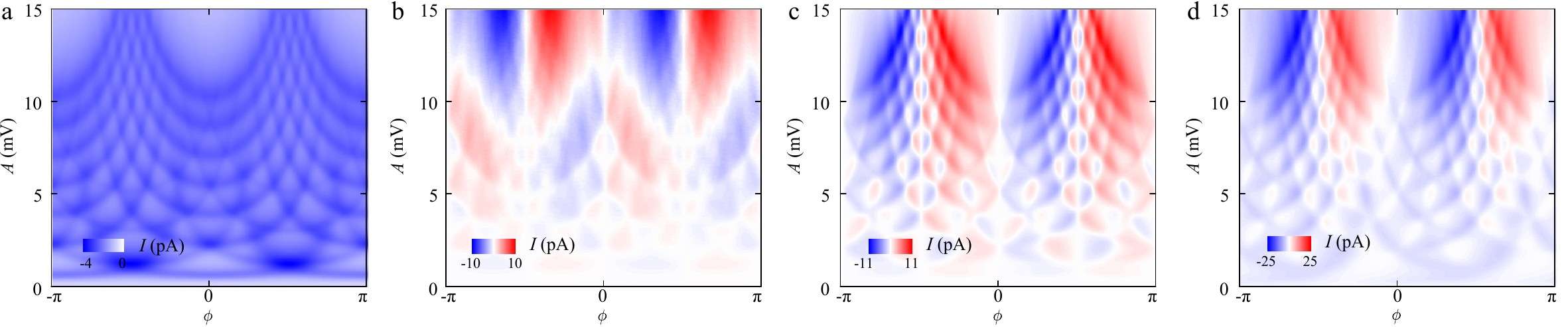}}
\caption{
Theoretical prediction of the current for $k=2$ shown in \protect\fig{fig2}{a}
of the main article,  $f_\text L=\,50$MHz $f_\text R=\,100$MHz.
The dc tunnel current $\bar G V$ (a), the pumped current $Q^\text{cycle}f$ (b), 
and the ac current (c) sum up to the total current plotted in (d).
}
\label{fig:k2}
\end{figure*}

The pumped charge is essentially determined by the scattering phase in the
prefactor of $S_n$.  Assuming again that the $\bar\Gamma_{L,R}$ are
practically constant at the resonance, we find \cite{Kashcheyevs2004b}
\begin{equation}
\label{app:Qapprox}
Q_\text{L}^n \simeq
\frac{e\bargl}{\bargl+\bargr}
\mathop{\mathrm{sign}}(\dot\mu_n) .
\end{equation}
This means that for $\bargl\gg\bargr$, an electron enters from
or leaves to the left lead depending on the sign of $\dot\mu_n$.  In the
opposite limit, the right lead is relevant and $Q_\text L^\text{res}$ is much
smaller than the elementary charge.  This also emphasizes the role of the
parameter dependence of the tunnel rates:
In the derivation of Eq.~\eqref{app:Qapprox}, we assumed that the
$\glr$ stays constant when a level crosses the chemical potential
of a lead at time $t=t_0$.  However, in a complete pump cycle, the level
will cross the chemical potential as many times from above as it crosses
from below.  Thus if $\glr$ were constant, the net charge pumped
from the left lead to the dot in the whole cycle would vanish.  In turn, we
will observe significant pumping from the left lead to the right lead
if $\gl\gg\gr$ when the dot level is lowered, while
$\gl\ll\gr$ when the level is raised.

For a two-terminal device, the pumped currents entering the dot from the
left and from the right lead compensate each other so that
$Q_\text L^\text{cycle}+Q_\text R^\text{cycle}=0$.  Nevertheless, the vector fields
appearing in the integral in Eq.~\eqref{app:Qx} may differ by more than
their sign, because the integrands are relevant only up to a gradient field
which vanishes under the closed line integral.  Therefore it is
more convenient to discuss the integrand of the symmetrized expression for
the pumped charge:
$Q^\text{cycle} = \frac{1}{2} (Q_\text L^\text{cycle}-Q_\text R^\text{cycle})
= \oint_\mathcal{C} d\vec v\cdot a(\vec v)$ with the vector field
\begin{equation}
\label{app:vectorfield}
\vec a = \frac{i}{4\pi}
\sum_n \Big\{ \big(S_n \mathop{\mathrm{grad}}
S_n^\dagger\big)_\text{LL}\Big|_{\mu_\text L}
- \big(S_n \mathop{\mathrm{grad}}
S_n^\dagger\big)_\text{RR}\Big|_{\mu_\text R} \Big\} .
\end{equation}

\Fig{fig:I,ax}{b} shows the first component of $\vec a(\vec v)$
for the parameters of our quantum dot.  A significant contribution requires
that the resonant level lies close to the chemical potential of a lead.
For small bias, this is also a necessary condition for the emergence of a
conductance peak, see \fig{fig:I,ax}{a}. To make direct use of
Brouwer's formula, one would transform via Stokes'
theorem the line integral into a surface integral [see remark after
\eq{app:Qx}] and would end up with an integration over the curvature shown
in \fig{fig:I,ax}{c}.

\subsection{Lissajous pumping and symmetries}

The gate voltages applied in the experiment follow Lissajous curves in
parameter space defined as
\begin{equation}
\label{app:lissajous}
\vkphi
= \begin{pmatrix} \tvl(t)\\ \tvr(t) \end{pmatrix}
= A\begin{pmatrix} \cos(\Omega t-\phi)\\ \cos(k\Omega t) \end{pmatrix} ,
\end{equation}
where henceforth $k$ is referred to as order and $\phi$ is the phase shift
between the two components.  Driving the two barriers with different but
commensurate frequencies is a particular feature of our experiment. The
amplitude $A$ is equal at both barriers and is typically several $\ec^0$ in
our experiments.  A central question is that of the symmetry properties of
the dc conductance and the pumped charge as a function of
the modulation parameters.

We start our symmetry considerations by noticing that the Lissajous
figures obey $\vkphi=\vec v_{k,\phi+2\pi/k}(t+2\pi/k\Omega)$,
namely that a phase shift of
$2\pi/k$ is canceled by a time shift of $2\pi/k\Omega$.  This implies that
the dc current patterns, being averaged over time, as a function
of $\phi$ possess a $k$-fold symmetry in $\phi$.

A further symmetry property of the three contributions to the current in
\eq{app:IL3} follows from the behavior of $\bar G_{\alpha\beta}$,
$Q^\text{cycle}$, and $I^\text{rect}$ under time reversal.  It plays a
crucial role for adiabatic pumping \cite{Brouwer1998a} and ratchet effects.
Below, we find $2k$ symmetry points at which $\bar G_{\alpha\beta}$ is
symmetric, while $Q^\text{cycle}$ and $I^\text{rect}$ are anti-symmetric
under time reversal.  For a derivation, we determine for each Lissajous
curve $\vec v_{k,\phi}$ a time-reversed partner with phase $\phi'$
which must fulfill the condition
\begin{equation}
\label{app:symm-cond}
\vec v_{k,\phi}(t) = \vec v_{k,\phi'}(-t+t_0),
\end{equation}
for all times $t$. The inversion $t\to -t$ is thereby accompanied by a
time offset $t_0$, which is permitted by the time periodicity of the
integrands in Eqs.\ \eqref{app:Irect}, \eqref{app:Gbar}, and
\eqref{app:Qx}.  Inserting \eq{app:lissajous} into \eq{app:symm-cond}
yields $\Omega t_0 = 2\pi\ell/k$ and $\phi' = -\phi + 2\pi(\ell/k+\ell')$
where $\ell=0,1,\ldots,k-1$ and $\ell'=0,1$.  The special phases
$\phi_{\ell\ell'}=\pi(\ell/k+\ell')$ fulfill $\phi=\phi'$ (with $t_0$
given above), so that the original and the time-reversed Lissajous curves
[defined by \eq{app:lissajous}] lie on top of each other, while they evolve
in opposite direction in time. Thus, the phases $\phi_{\ell\ell'}$
define $2k$ in-equivalent points with TRS.
An interesting observation is that there, the Lissajous curve takes the
form $\tvr(\tvl) = \pm\cos[k\arccos(\pm \tvl)]$. This defines the $k$th
Chebyshev polynomial \cite{Gradshteyn1994a} which represents a degenerate
loop that does not enclose a finite area. As a consequence, $Q^\text{cycle}$ and $I^\text{rect}$ (with odd symmetry under time-reversal) vanish. Hence,  at the symmetry points the current takes the value $\bar G V$ and, in particular, vanishes for $V=0$.

Substituting in \eq{app:lissajous} the phase $\phi$ by the deviation from
the symmetry point, $\Delta\phi = \phi-\phi_{\ell\ell'}$, one can see that
time reversal corresponds to $\Delta\phi\to-\Delta\phi$.  Thus, the
behavior of the integrals in Eqs.~\eqref{app:Irect}, \eqref{app:Gbar}, and
\eqref{app:Qx} under time reversal allows us to draw conclusions about the
symmetries of $\bar G_{\alpha\beta}$ and $Q^\text{cycle}$ as a function of
$\Delta\phi$.  To be specific: Since the average conductance \eq{app:Gbar}
is invariant under time reversal, it must be an even function of
$\Delta\phi$.  By contrast, the charge pumped per cycle, \eq{app:Qx},
changes its sign upon time reversal.  Consequently, it must be an odd
function of $\Delta\phi$.  In short, despite the only $k$-fold symmetry of
the Lissajous curves \eq{app:lissajous}, we find $2k$ in-equivalent phases
at which $\bar G_{\alpha\beta}$ is symmetric, while $Q^\text{cycle}$ is
anti-symmetric.

Our experimental results indicate that the rf modulation of gate voltages
induces a tiny ac bias. Its origin is a tiny capacitive coupling
between the modulated gates and the 2D leads of the QD. Therefore, we model
it as modulation of the chemical potentials $\tilde\mu_\alpha(t) =
\mu_\alpha + ew_\alpha(t)$ with a phase shift of $-\pi/2$ compared to the
modulation of the gate voltages [described in Eq.~(1) of the main article]:
\begin{equation}
\label{app:lissajous_ac}
\vec w_{k,\phi}(t) =
\begin{pmatrix} w_\text{L}(t)\\ w_\text{R}(t) \end{pmatrix}
= \kappa_\text{ac}A\begin{pmatrix} \sin(\Omega t-\phi)\\ \sin(k\Omega t) \end{pmatrix} .
\end{equation}
The capacitive coupling constant $\kappa_\text{ac}$ turns out the be of the
order $5\cdot 10^{-4}$ and is expected to decrease with the modulation
frequency, $\kappa_\text{ac}\propto\Omega^{-1}$.

In comparison to the gate voltages in \eq{app:lissajous}, the ac
modulation of the chemical potentials, $\vec w$, contains a sine instead of
a cosine. In a symmetry analysis along the lines above, this finally leads
to a minus sign.  Therefore the integrad in Eq.~\eqref{app:Irect} is an odd
function of $t$ (besides a time shift by $t_0$), so that the rectified
current changes its sign under time reversal.  Consequently,
$I^\text{rect}$ is anti-symmetric in $\Delta\phi$.

In Figs.~\ref{fig:k1} and \ref{fig:k2}, we show for $k=1,2$ the three contribution to
the current in \eq{current} of the main text, i.e., the dc current $\bar GV$,
the pump current $Q^\text{cycle}f$, and the contribution of the rectified
ac bias, $I^\text{rect}$.  This visualizes the symmetry of the dc current
and the anti-symmetry of the second and the third contribution. Notably,
for the case of a tiny dc voltage of $V=1\mu$V considered here, the overall
behavior [panel (d)] is dominated by the anti-symmetric contributions.

\section{Additional data}

\begin{figure*}[ht]
\includegraphics[scale=1]{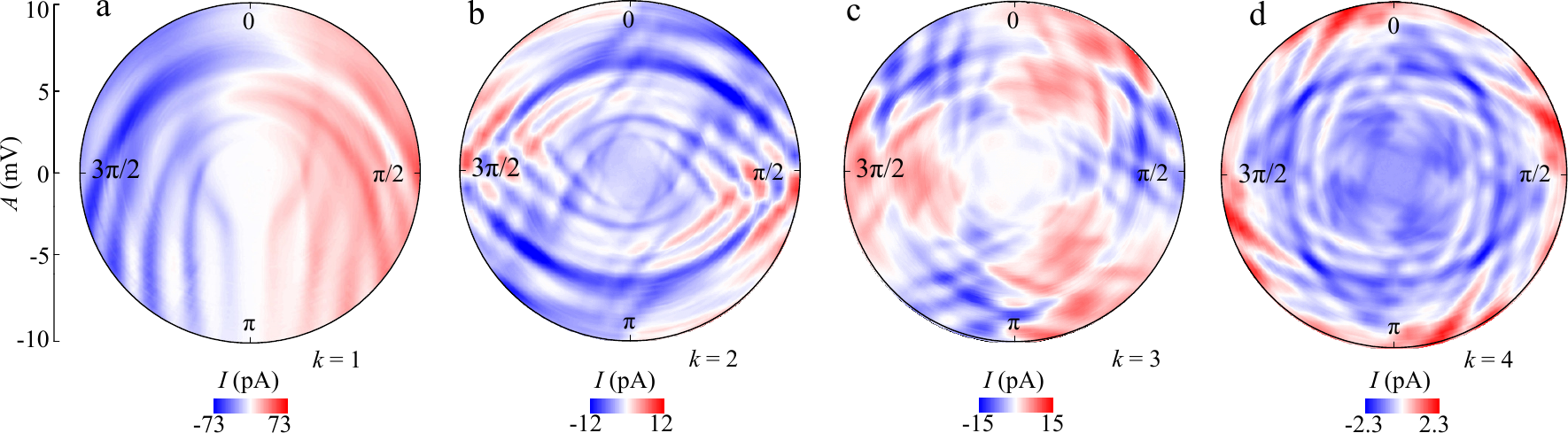}
\caption{\label{fig:dk1}
Measured current $I$ versus modulation amplitude and phase
for (a) $k=1$ and $f_\text L=f_\text R=100\,$MHz, (b) $k=2$ and $f_\text L=\,100$MHz $f_\text R=\,200$MHz, (c) $k=3$ and $f_\text L=\,50$MHz $f_\text R=\,150$MHz, (d) $k=4$ and $f_\text L=\,50$MHz $f_\text R=\,200$MHz. The data for $k=3$ are identical to those in Fig.\ \ref{fig2} of the main article. The working point is defined by $\vl=\,-202$mV and $\vr=\,-91$mV as marked in Fig.\ \ref{fig1}(a) of the main article.}
\end{figure*}
\begin{figure*}[ht]
\includegraphics[scale=1]{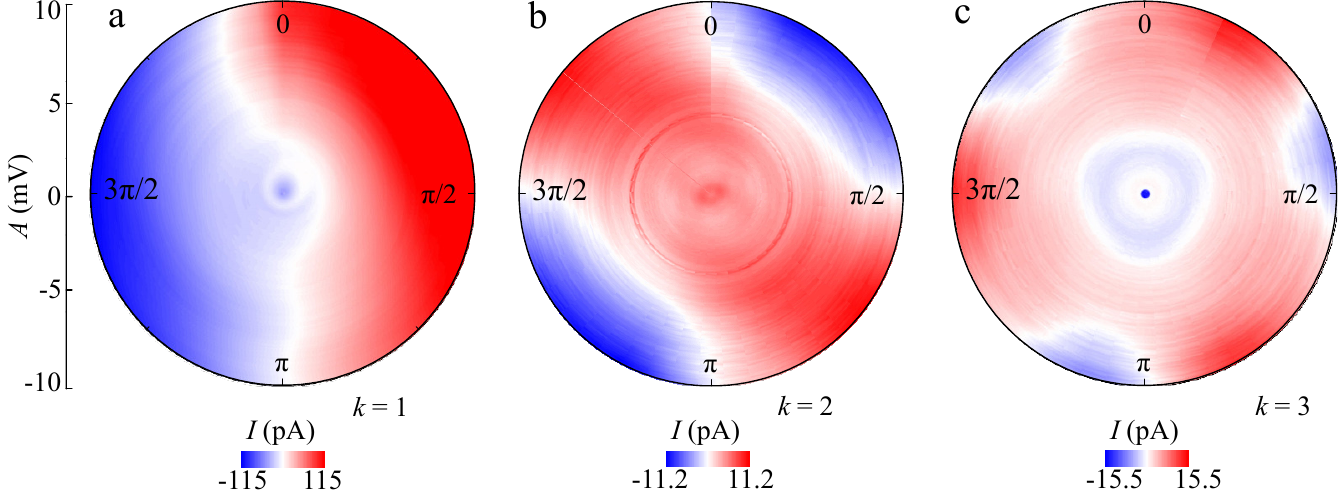} 
\caption{\label{fig:dk2}
Measured current $I$ versus modulation amplitude and phase for  (a) $k=1$ and $f_\text L=f_\text R=100\,$MHz, (b) $k=2$ and $f_\text L=50\,$MHz $f_\text R=\,100$MHz, (c) $k=3$ and $f_\text L=50\,$MHz $f_\text R=\,150$MHz. The working point is defined by $\tvl=\,12.5$mV and $\tvr=\,12.5$mV near the upper right corner of Fig.\ \ref{fig1}(a) of the main article. The enhanced dot-lead tunnel coupling washes out most fine structures compared to the data in Fig.\ \ref{fig:dk1} above.
}
\end{figure*}

In \fig{fig:dk1}{} we compare the measured Lissajous ratchet data for $k=1,2,3$ also discussed in the main article but all in spherical coordinates and add another data set for $k=4$. The data contain the predicted $k+1$-fold symmetry but also show a complex detailed structure caused by the combination of the three contributions in \eq{current} of the main article, namely the time averaged conductance $\bar GV$, the ratchet current $Q^\text{cycle}f$ and $I^\text{rect}$. It is possible to reduce the information in similar measurements, such that the $k+1$-fold symmetry is more evident, by increasing the dot-lead tunnel couplings. We do this by shifting the working point to the mixed valance regime where the Coulomb current peaks overlap strongly ($\tvl=\,12.5$mV and $\tvr=\,12.5$mV, see \fig{fig1}{a} of the main article). The result is a strong broadening which thoroughly washes out most fine structure of the current. In \fig{fig:dk2}{}
we present such data up to $k=3$. These measurement are a clear demonstration of the Lissajous ratchet effect as they emphasize the symmetry properties of $Q^\text{cycle}$.

\section{Applications and alternative realizations}

In our Lissajous rocking ratchet a combination of rf excitations breaks the time reversal symmetry and yields a dc current.  It represents a new method to access radio frequency information by mutually comparing the frequencies, amplitudes and phases of  rf signals. As such, Lissajous rocking ratchets could find applications as rf comparator or serve as detector or filter components. Our quantum dot implementation is interesting for quantum information processing applications, as it allows one to compare and process small rf-signal on-chip. In a simplified picture a Lissajous ratchet resembles an on-chip rf Lock-In amplifier: to illustrate this idea imagine that we apply a clean rf reference to one gate and a noisy signal to the second gate. The resulting dc current strongly depends on the relative phase, amplitude and frequency of the two signals. By measuring a time averaged dc current, our device filters the time periodic carrier signal while averaging out uncorrelated noise, just as a Lock-In amplifier does. However, the Lissajous ratchet allows a detailed comparison of mutual rf-signals going well beyond the scope of a Lock-In amplifier.

\begin{figure}[ht]
\includegraphics[width=1\columnwidth]{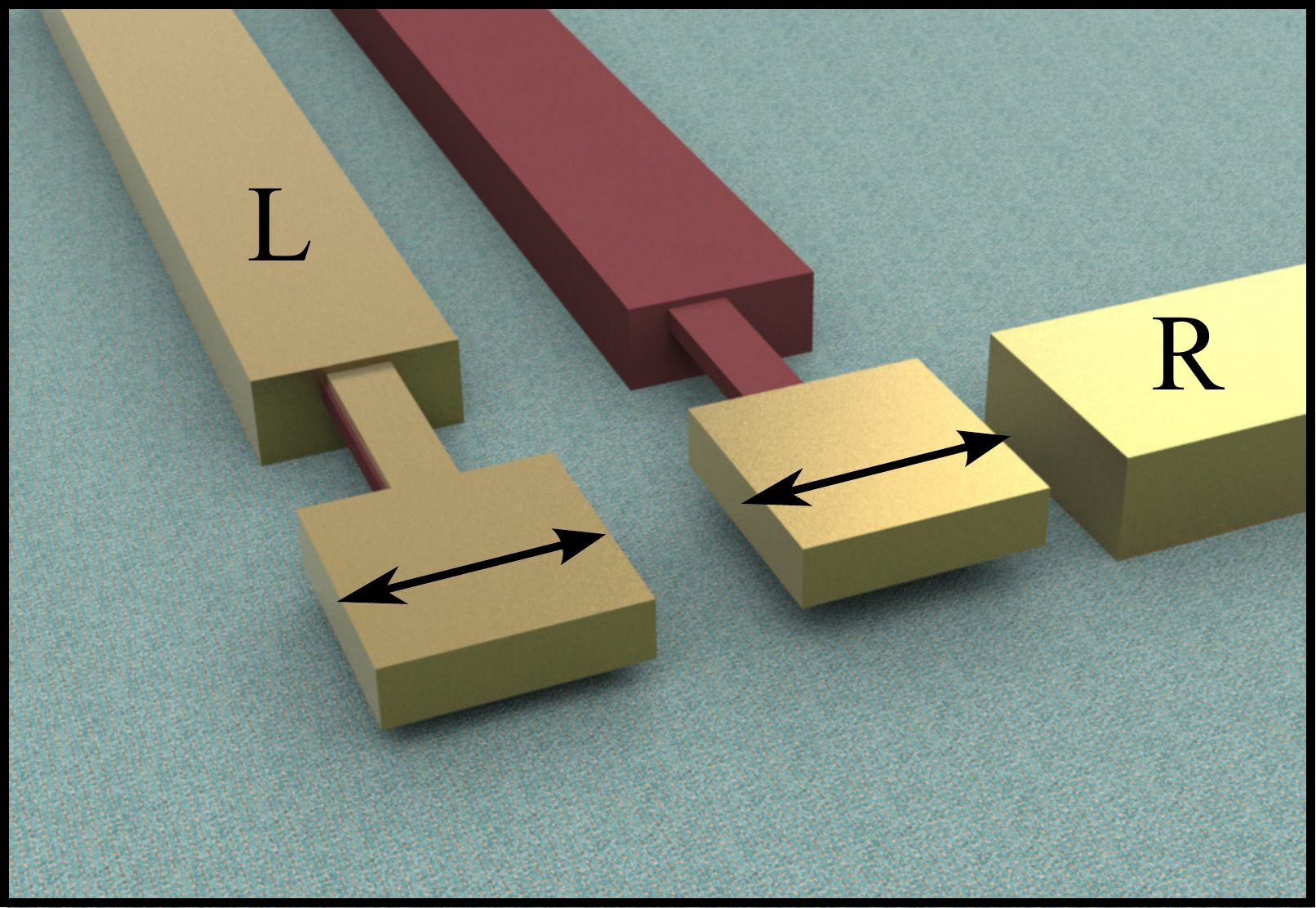}
\caption{\label{fig:cradle}
Nanomechanical Newton cradle discussed in the main text.
}
\end{figure}
Lissajous rocking ratchets are not limited to QD circuits but could be realized in a variety of systems, for instance in macroscopic electronics or mechanics or in superconducting circuits. In the following we propose a possible realization in a  nanoelectromechanical systems. In \fig{fig:cradle}{} we extend the already realized ``nano bell'' \cite{Koenig2008} and illustrate what we mean by ``nanomechanical Newton cradle''. The device is composed of two cantilevers and a fixed metallic contact (R). The central cantilever is patterned with an isolated metallic island while the left one contains an island but electrically connected to a second lead (L). The islands are positioned such, that the central island can touch both the left and right leads whenever both cantilevers are vibrating with sufficiently high amplitude. A voltage can be applied between L and R and current can be measured. We assume that both cantilevers can be externally driven (at predefined relative phases) at their mutual eigen frequencies and that the latter are tunable, \eg, via capacitive coupling to additional gates or by using dielectric forces \cite{Unterreithmeier2009}. We also assume that the central island is small enough to allow for single electron transport based on Coulomb blockade, namely that we can treat it as a QD \cite{Koenig2008}. The electronic levels of the islands are then modulated by the same capacitive coupling that is used to mechanically drive the cantilevers. In such a device the coupling between the islands and that between contact R and the center island are strongly time dependent, zero most of the time and strong during touching. This system is highly tunable and very close in spirit to our QD circuit: importantly, the time dependence of the couplings and island levels are linked, a central precondition to define a Lissajous ratchet.

\bibliography{literature,supplement}

\end{document}